 \documentclass[final,5p,times,twocolumn]{elsarticle}

\usepackage{amssymb}
\usepackage{caption}
\usepackage{graphicx}
\usepackage{epstopdf, epsfig}
\usepackage{amsmath}

\usepackage{xcolor}
\usepackage{wasysym} 
\usepackage{dblfloatfix}
\usepackage[normalem]{ulem}
\usepackage[nameinlink,noabbrev]{cleveref}
\journal{Experimental Thermal and Fluid Science}

\usepackage{makecell}
\usepackage{tikz}
\usepackage{booktabs}
\usepackage{multirow}
\usepackage{siunitx}
\usepackage{array}
\newcommand{\Um}{U_m}
\newcommand{\Palign}{P_{\mathrm{align}}}
\newcommand{\Pdiv}{P_{\mathrm{div}}}
\newcommand{\Ptrace}{P_{\mathrm{trace}}}
\newcommand{\Pzz}{P_{zz}}
\newcommand{\Dtd}{D_{2D}}
\newcommand{\ARS}{A_{RS}}

\usepackage{etoolbox}

\makeatletter

\patchcmd{\elsarticleprelims}
  {\unvbox\elsarticlegrabsbox}
  {\unvbox\elsarticlegrabsbox\vfill}
  {}{}

\patchcmd{\elsarticleprelims}
  {\unvbox\elsarticlehighlightsbox}
  {\unvbox\elsarticlehighlightsbox\vfill}
  {}{}

\makeatother

\begin{document}

\begin{frontmatter}

\title{Flow separation and turbulence production over trapezoidal protrusions}

\author[inst1]{Shahar Zuckerman}
\affiliation[inst1]{organization={Faculty of Mechanical Engineering},
 addressline={Technion - Israel Institute of Technology}, 
 city={Haifa},
 postcode={3200003}, 
 country={Israel}}

\author[inst1]{Jibu Tom Jose}
\author[inst1]{Ori Levi}
\author[inst1]{Omri Ram}



\begin{abstract}
The influence of rear-face inclination on turbulence around a surface-mounted protrusion depends on the separated flow approaching its trailing edge. This dependence is examined experimentally using stereoscopic particle image velocimetry over twelve trapezoidal protrusions with upper-surface length-to-height ratios $L/h=1,2,3,4$ and rear-face angles of $30^\circ$, $45^\circ$, and $90^\circ$. Measurements were conducted at a height-based Reynolds number of approximately $5\times10^4$, with an incoming boundary-layer thickness comparable to the protrusion height. For $L/h=1$ and 2, a connected reverse-flow region extends over the upper surface into the wake. For $L/h=3$ and 4, upper-surface reattachment separates the two regions. Length governs this change in topology and the variation in turbulence levels at common spatial locations, whereas rear-face inclination has its strongest influence close to the surface. From $L/h=2$ onward, shallower faces have higher near-face turbulent kinetic energy, while vertical faces have lower energy but a larger spanwise fraction. A decomposition of the in-plane deviatoric production separates the effects of Reynolds-stress and mean-strain magnitudes from those of their relative principal-axis orientation. At the inclined faces, this contribution changes from negative to positive between $L/h=1$ and 2, before upper-surface reattachment occurs. Its subsequent increase with length is driven mainly by increasing tensor magnitudes, with the largest contribution at $L/h=4$ and $30^\circ$, where strong production remains close to the rear face. These results show that local turbulence production depends jointly on the flow state approaching the trailing edge and the position of the energetic shear layer relative to the rear surface.
\end{abstract}




\end{frontmatter}
\begin{table*}[!t]
\centering
\begingroup
\footnotesize
\setlength{\fboxsep}{7pt}
\setlength{\fboxrule}{0.4pt}
\renewcommand{\arraystretch}{1.12}

\fbox{%
\begin{minipage}{\dimexpr\textwidth-2\fboxsep-2\fboxrule\relax}
\textbf{Nomenclature}
\par\vspace{6pt}

\noindent
\begin{minipage}[t]{0.485\linewidth}
\vspace{0pt}
\begin{tabular}{@{}
  >{\raggedright\arraybackslash}p{0.23\linewidth}
  @{\hspace{0.025\linewidth}}
  >{\raggedright\arraybackslash}p{0.745\linewidth}
  @{}}

\multicolumn{2}{@{}l}{\textit{Symbols }}\\[3pt]

$A_{RS}$
& Band-averaged stress--strain amplitude,
  $2\overline{r_Rr_S}_{\mathrm{TE}}$
  [$\mathrm{m^2\,s^{-3}}$] \\

$B$
& Protrusion span [$\mathrm{m}$] \\

$C_R$
& Half-trace of the in-plane Reynolds-stress tensor
  [$\mathrm{m^2\,s^{-2}}$] \\

$C_S$
& Half-trace of the in-plane mean-strain tensor
  [$\mathrm{s^{-1}}$] \\

$D_{2D}$
& In-plane mean-velocity divergence,
  $\partial U/\partial x+\partial V/\partial y$
  [$\mathrm{s^{-1}}$] \\

$f_z$
& Spanwise energy fraction, $R_{zz}/(2k)$ [$-$] \\

$F_{>45}$
& Amplitude-weighted fraction with
  $\Delta\theta>45^\circ$ [$\%$] \\

$h$
& Protrusion height [$\mathrm{m}$] \\

$\mathbf{I}$
& Identity tensor [$-$] \\

$k$
& Turbulent kinetic energy per unit mass,
  $R_{ii}/2$ [$\mathrm{m^2\,s^{-2}}$] \\

$L$
& Upper-surface length [$\mathrm{m}$] \\

$L^*$
& Normalized upper-surface length, $L/h$ [$-$] \\

$L_{\mathrm{RF}}^*$
& Streamwise extent of a connected reverse-flow
  region, normalized by $h$ [$-$] \\

$O_{\mathrm{TE}}$
& Upper trailing-edge corner, $(L,h)$ \\

$P$
& Turbulent kinetic-energy production per unit mass
  [$\mathrm{m^2\,s^{-3}}$] \\

$P_{2D}$
& In-plane production,
  $-\mathbf{R}_{2D}:\mathbf{S}_{2D}$
  [$\mathrm{m^2\,s^{-3}}$] \\

$P_{\mathrm{align}}$
& Deviatoric contribution to in-plane production
  [$\mathrm{m^2\,s^{-3}}$] \\

$P_{\mathrm{div}}$
& Divergence-associated contribution,
  $P_{\mathrm{trace}}+P_{zz}$
  [$\mathrm{m^2\,s^{-3}}$] \\

$P_{\mathrm{trace}}$
& Isotropic contribution to in-plane production
  [$\mathrm{m^2\,s^{-3}}$] \\

$P_{zz}$
& Spanwise normal-stress contribution to production
  [$\mathrm{m^2\,s^{-3}}$] \\

$P_{\mathrm{res}}$
& Measurable production contribution that vanishes
  under exact midplane symmetry
  [$\mathrm{m^2\,s^{-3}}$] \\

$\mathbf{Q}$
& Rotation matrix from $(x,y)$ to $(t,n)$ [$-$] \\

$q_{zz}$
& Normalized spanwise variance, $R_{zz}/U_m^2$ [$-$] \\

$R_{ij}$
& Reynolds stresses,
  $\langle u_i'u_j'\rangle$
  [$\mathrm{m^2\,s^{-2}}$] \\

$R_{zz,\mathrm{floor}}^*$
& Estimated normalized background level of
  spanwise variance [$-$] \\

$\mathbf{R}'$
& Deviatoric part of $\mathbf{R}_{2D}$
  [$\mathrm{m^2\,s^{-2}}$] \\

$Re_h$
& Height-based Reynolds number, $U_mh/\nu$ [$-$] \\

$r_R$
& Deviatoric stress magnitude
  [$\mathrm{m^2\,s^{-2}}$] \\

$r_S$
& Deviatoric strain-rate magnitude
  [$\mathrm{s^{-1}}$] \\

$S_{ij}$
& Mean strain-rate components
  [$\mathrm{s^{-1}}$] \\

$\mathbf{S}'$
& Deviatoric part of $\mathbf{S}_{2D}$
  [$\mathrm{s^{-1}}$] \\

$s$
& Arclength along the rear face, measured from
  the upper TE corner toward the wall [$\mathrm{m}$] \\

$SS$
& Sum of squares in the geometric decomposition \\

$\mathrm{St}$
& Particle Stokes number [$-$] \\

\end{tabular}
\end{minipage}
\hfill
\begin{minipage}[t]{0.485\linewidth}
\vspace{0pt}
\begin{tabular}{@{}
  >{\raggedright\arraybackslash}p{0.23\linewidth}
  @{\hspace{0.025\linewidth}}
  >{\raggedright\arraybackslash}p{0.745\linewidth}
  @{}}

$u,v,w$
& Instantaneous streamwise, wall-normal and
  spanwise velocities [$\mathrm{m\,s^{-1}}$] \\

$U,V,W$
& Ensemble-mean velocity components
  [$\mathrm{m\,s^{-1}}$] \\

$u',v',w'$
& Velocity fluctuations
  [$\mathrm{m\,s^{-1}}$] \\

$U_m$
& Tunnel reference velocity [$\mathrm{m\,s^{-1}}$] \\

$x,y,z$
& Streamwise, wall-normal and spanwise coordinates
  [$\mathrm{m}$] \\

$x_r$
& Global streamwise reattachment coordinate
  [$\mathrm{m}$] \\

$X_r^*$
& Normalized reattachment coordinate, $x_r/h$ [$-$] \\

$Y_{ij}$
& Scalar response for length case $i$ and
  angle case $j$ \\[5pt]

\multicolumn{2}{@{}l}{\textit{Greek symbols}}\\[3pt]

$\alpha$
& Rear-face angle measured from the wall [$^\circ$] \\

$\delta_{99}$
& Incoming boundary-layer thickness [$\mathrm{m}$] \\

$\Delta$
& PIV vector spacing [$\mathrm{m}$] \\

$\Delta\theta$
& Local misalignment of the stress and strain
  principal axes, $0^\circ$--$90^\circ$ \\

$\eta_{\mathrm{sym}}$
& Relative measurable symmetry residual,
  $\overline{|P_{\mathrm{res}}|}/\overline{|P|}$ [$-$] \\

$\theta_A$
& Direction of the area-averaged TE velocity
  vector, measured from the positive $x$ direction
  [$^\circ$] \\

$\theta_R,\theta_S$
& Principal directions of stress and strain,
  measured from $t$ toward $n$ [$^\circ$] \\

$\nu$
& Kinematic viscosity [$\mathrm{m^2\,s^{-1}}$] \\

$\nu_t$
& Kinematic eddy viscosity [$\mathrm{m^2\,s^{-1}}$] \\

$\rho_p$
& Tracer-particle density [$\mathrm{kg\,m^{-3}}$] \\

$\sigma_{\mathrm{MAD}}$
& Robust scale estimate,
  $1.4826\,\operatorname{MAD}(q_{zz})$ [$-$] \\[5pt]

\multicolumn{2}{@{}l}{\textit{Subscripts and notation}}\\[3pt]

$2D$
& In-plane quantity in the $(x,y)$ plane \\

$A$
& Average over the valid TE-corner sampling area \\

$RS$
& Weighting by the local amplitude $r_Rr_S$ \\

$\mathrm{TE}$
& Average over the trailing-edge-face band \\

$t,n$
& Face-tangential and outward-normal directions \\

$\langle\cdot\rangle$
& Ensemble average \\

$\overline{(\cdot)}$
& Spatial average over the specified region \\

$\widetilde{\Delta\theta}$
& Amplitude-weighted median misalignment \\

$(\cdot)^*$
& Nondimensional quantity; lengths are scaled by
  $h$ and Reynolds stresses by $U_m^2$ \\

$(\cdot)'$
& Velocity fluctuation for a velocity component;
  deviatoric part for an in-plane tensor \\

$:$ 
& Double tensor contraction \\[5pt]

\end{tabular}
\end{minipage}

\end{minipage}%
}
\endgroup
\end{table*}

\section{Introduction}
\label{sec:intro}
\label{sec:problem}

Surfaces in engineering and natural flows are seldom smooth. Rivets, weld beads, sensors, and erosion-induced roughness disrupt the boundary layer and can promote local separation, increasing drag and turbulence production \citep{castro1977flow}. In aerospace applications, discrete roughness elements can trigger laminar-to-turbulent transition and add parasitic drag \citep{choudhari2010llaminar}, while in turbomachinery, manufacturing imperfections and operational wear degrade aerodynamic performance \citep{leipold2000influence}. In thermal management, separation and reattachment are deliberately exploited through ribs and protrusions that promote mixing and enhance convective heat transfer \citep{zhang2021turbulent,meinders1997application,acharya1998heat}. In marine and hydraulic systems, surface defects can promote cavitation \citep{brennen2014cavitation,Arndt2002}, where inception in the separated shear layer is influenced by turbulence-related pressure fluctuations \citep{agarwal2023pressure}. In natural flows, boulders and coarse sediments shape near-bed turbulence and sediment entrainment \citep{Dey2014}, while vegetation, buildings and terrain features modify the exchange of momentum and scalars within the atmospheric boundary layer \citep{stull2012introduction}.

Boundary-layer separation is a principal mechanism through which surface features affect aerodynamic and thermal performance. Abrupt changes in geometry can cause the flow to detach, leaving a recirculating region bounded by a separated shear layer \citep{Simpson1989}. Such flows exhibit elevated turbulence intensity and pressure fluctuations and can contribute to increased drag, structural vibration and flow-induced noise \citep{Simpson1989,agarwal2023pressure}. Time-resolved measurements have identified vortex shedding, shear-layer flapping and intermittent reattachment as sources of broadband fluctuations \citep{Cherry1984,kiya1985structure,shi2010influence,liu2008unsteady,zhang2017separated}. The exchange of energy between the mean flow and the turbulent fluctuations is quantified by turbulent kinetic-energy production, defined as the negative contraction of the Reynolds-stress tensor with the mean velocity gradient \citep{pope2000turbulent}. 

\begin{figure*}[t]
 \centering
 \includegraphics[width=0.75\textwidth]{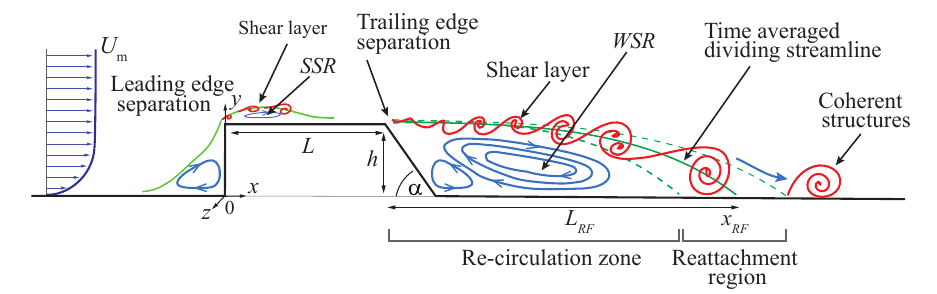}
 \caption{Schematic of the flow over a surface-mounted trapezoidal
 protrusion. The sketch shows the principal separated-flow structures in a generic configuration. The connectivity of the upper-surface and wake recirculation regions depends on $L^*$.}
 \label{fig:schematic}
\end{figure*}

Surface-mounted obstacles with sharp leading edges, such as ribs, steps, and prisms, are canonical configurations for this problem because their sharp leading edges fix the separation point \citep{kiya1983structure,Martinuzzi1993}. Figure~\ref{fig:schematic} sketches the principal flow regions of a surface-mounted protrusion. The separated flow can include an upstream bubble ahead of the forward face, a recirculation region over the upper surface, and a downstream wake bubble. Their development and interaction depend strongly on the streamwise length $L$ of the body. For long protrusions, the shear layer leaving the leading edge reattaches on the upper surface, leaving distinct upper-surface and wake recirculation regions. As the body length increases, these regions become largely independent and downstream turbulence levels decrease \citep{Antoniou1988,van2018effect}. For short protrusions, the leading-edge shear layer does not reattach before reaching the rear face, so the upper-surface region and the wake form a single elongated recirculation region. At intermediate lengths, reattachment can be partial and intermittent, accompanied by shear-layer flapping \citep{van2018effect,Fang2019}. Early measurements over rectangular obstacles showed that the wake reattachment distance decreases as the body length increases \citep{bergeles1983flow,Antoniou1988}. The body length required for the leading-edge separation to reattach on the upper surface also depends on the incoming flow. Studies of backward-facing steps have shown that Reynolds number and boundary-layer state affect reattachment \citep{Armaly1983}, while measurements over forward-facing steps have linked reattachment distance to mixing and the momentum deficit in the approaching layer \citep{Sherry2010,GRAZIANI2017220}. Large-scale structures in that layer also affect unsteadiness. Low-momentum regions have been linked to intermittent separation upstream of a forward-facing step \citep{Pearson2013}, and alternating low- and high-velocity streaks to flapping of the upper-surface bubble over a forward-backward-facing step \citep{Fang2019}. In the present experiments, the incoming boundary layer is approximately as thick as the obstacle. The incoming flow conditions were held constant across all configurations, so their effects on separation and reattachment were not examined in this study.

In addition to body length, the inclination of the trailing-edge (TE) face can affect the separated flow. Evidence comes from several related configurations. Using laser Doppler measurements in confined channel flow, \citet{ruck1993flow} investigated backward-facing steps with face inclinations between $10^\circ$ and $90^\circ$. They found that the reattachment length changed little as the inclination decreased from $90^\circ$ to $45^\circ$, but shortened substantially at smaller angles. On smooth ramps, separation is not fixed by an edge, and the development of the separated shear layer depends on the imposed adverse pressure gradient \citep{simmons2024experimental,songEaton2004}. In a configuration more closely related to the present geometry, \citet{ali2013flow} used particle image velocimetry (PIV) and liquid crystal thermography to examine flow and heat transfer behind trapezoidal ribs. They found that changing the chamfering angle modifies vortical structures near the leeward corner and affects local and downstream heat transfer. Measurements around the Ahmed body also showed that the rear slant angle changes the locations of critical points in the mean wake and the dominant frequencies of its unsteadiness \citep{tunay2014effects}. On elongated bluff bodies, the state of the leading-edge separated layer arriving at the trailing edge influences trailing-edge vortex formation and shedding \citep{duan2020effects}. Similar geometric effects occur over dunes, where numerical simulations showed that the mean lee-side angle governs the occurrence of separation and the strength of the turbulent wake, with additional effects from the maximum local slope and its position \citep{lefebvre2023influence}. Axisymmetric expansions provide a more direct link between wall inclination and turbulence production. Comparisons of abrupt and gradual expansions attributed differences in production to changes in the return flow, which ascends the sloped wall toward the outer stream in the gradual case but is disrupted by a secondary corner vortex in the abrupt case \citep{jose2026effect,jose2026coherent}.

These studies show that the inclination of the rear face affects the path of the separated shear layer and the recirculating flow beneath it. Its effect on turbulence close to the rear face and the dependence of this effect on body length remain less clear. For long protrusions, the leading-edge shear layer reattaches on the upper surface, allowing forward flow to approach the trailing edge. For short protrusions, the trailing edge remains within the recirculation region extending from the leading edge into the wake. The local flow near a rear face of a given inclination therefore differs between the two cases. To examine this dependence, we first characterize the velocity field near the trailing edge for different protrusion lengths and trailing-edge angles. We then resolve the Reynolds stresses in coordinates aligned with the rear face to determine the distribution of turbulent kinetic energy among the face-tangential, face-normal, and spanwise velocity fluctuations. Stereoscopic particle image velocimetry (SPIV) provides all three velocity components needed for this analysis. We then examine how the Reynolds stresses and mean strain rates govern energy transfer between the mean flow and the turbulent fluctuations. A scalar production field identifies where energy is transferred between the mean flow and the fluctuations \citep{jose2026effect}, but it does not distinguish the roles of tensor magnitude and relative orientation. We therefore decompose the in-plane production using the isotropic and deviatoric parts of the Reynolds-stress and mean-strain tensors. The contribution from the deviatoric parts separates exactly into an amplitude, determined by their magnitudes, and an orientation factor, determined by the angle between their principal axes. This orientation factor sets the sign of the deviatoric contribution. The relative orientation of the stress and strain tensors is commonly used to assess the coaxiality assumed by eddy-viscosity closures \citep{pope1975effective,schmitt2007boussinesq,lengani2020principal,corsini2025transport}. Here, we use it together with the tensor magnitudes to determine how body length and rear-face inclination affect local turbulence production.\\

\section{Experimental Methodology}
\label{sec:methods}

\subsection{Facility and Flow Conditions}
\label{sec:facility}
\begin{figure}[t]
\centering
\includegraphics[width=0.45\textwidth]{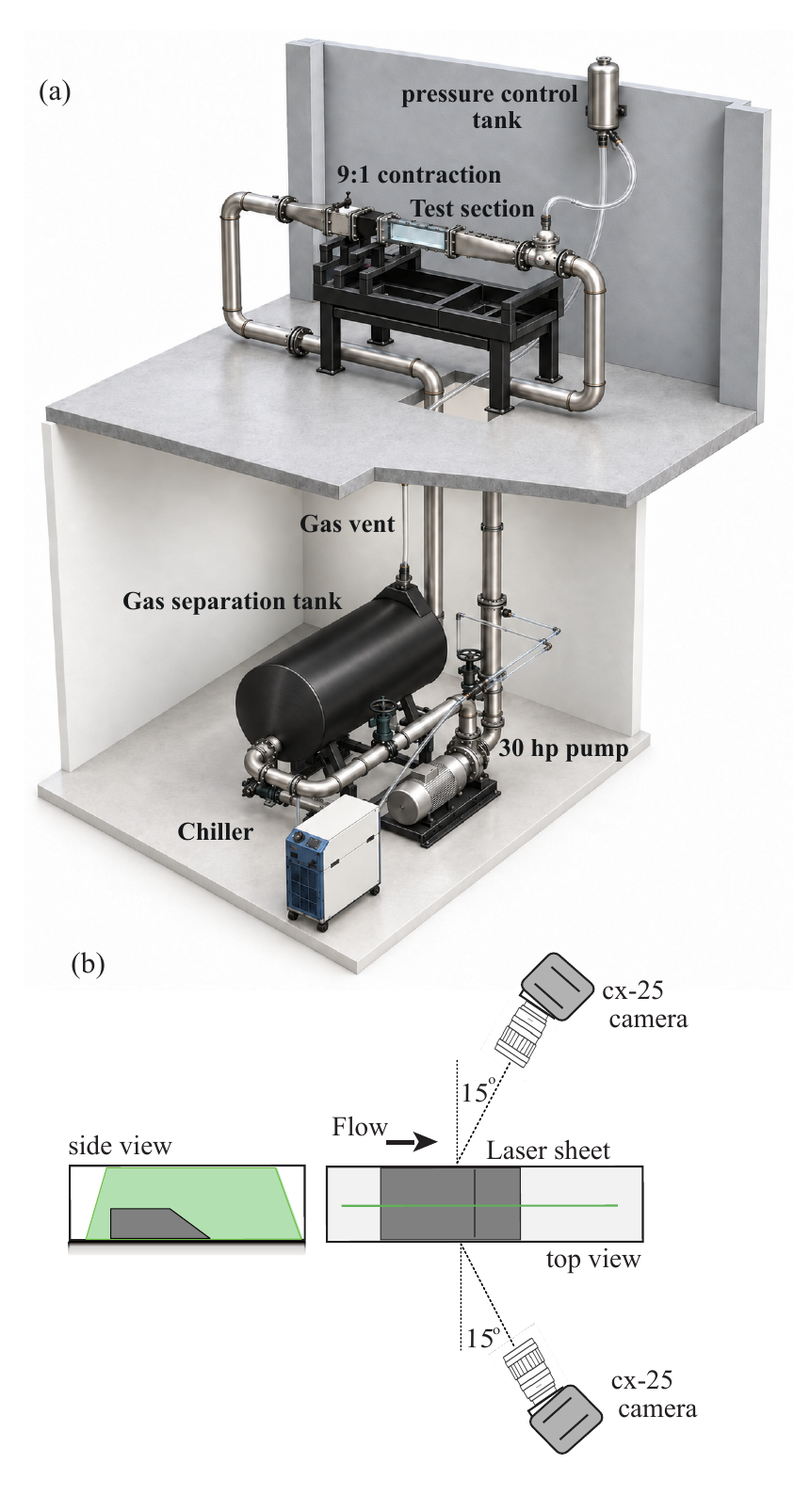}
\caption{(a) Closed-loop water-tunnel facility. (b) Schematic of the mounted
protrusion and stereoscopic PIV arrangement in the test section.}
\label{expsys}
\end{figure}

Experiments were conducted in the closed-loop water tunnel of the Transient Fluid Mechanics Laboratory at Technion--Israel Institute of Technology, shown in Fig.~\ref{expsys}(a). The transparent acrylic test section is $25~\mathrm{mm}$ wide and $200~\mathrm{mm}$ high, providing optical access for illumination and imaging. A $30~\mathrm{hp}$ centrifugal pump, controlled by a variable-frequency drive, allows continuous adjustment of the free-stream velocity up to $15~\mathrm{m\,s^{-1}}$. The tunnel, pump and test section are mounted on a heavy, vibration-damped frame to minimize mechanical disturbance of the measurements.

The models are two-dimensional trapezoidal protrusions machined from acrylic (PMMA) on a CNC mill and then polished. Each model spans the full test-section width and has a height of $h=10~\mathrm{mm}$ and a vertical forward face. The sharp upper corner fixes the leading-edge separation point. The models differ only in their upper-surface length $L$ and rear-face inclination $\alpha$, measured from the wall. Four normalized lengths, $L^*=L/h=1$, 2, 3 and 4, were combined with three angles, $\alpha=30^\circ$, $45^\circ$ and $90^\circ$, giving twelve configurations.

All experiments were conducted in water at approximately $20~^{\circ}\mathrm{C}$ and a nominal tunnel speed of $5~\mathrm{m\,s^{-1}}$, with no cavitation observed on the models. The measured tunnel reference speed is denoted $\Um$, giving a height-based Reynolds number of $Re_h=\Um h/\nu\simeq5\times10^4$, where $\nu$ is the kinematic viscosity. Measurements of the undisturbed boundary layer guided the choice of model location, where $\delta_{99}\simeq10~\mathrm{mm}$, approximately equal to the protrusion height. The outer-flow streamwise turbulence intensity was approximately $3.2\%$. The incoming flow conditions were held constant across all configurations.

The water was seeded with silver-coated hollow glass spheres with a diameter of $13~\mu\mathrm{m}$ and a density of $\rho_p=1600~\mathrm{kg\,m^{-3}}$. The corresponding particle Stokes number is $\mathrm{St}\sim10^{-2}\ll1$, indicating that the tracers faithfully follow the flow at the scales resolved by the measurements.

\subsection{Stereoscopic PIV and Data Processing}
\label{sec:piv}
Three-component velocity fields were measured using the stereoscopic PIV arrangement shown in Fig.~\ref{expsys}(b). Two LaVision Imager CX-25 cameras, each equipped with a $100~\mathrm{mm}$ macro lens and a Scheimpflug adapter, viewed the measurement plane at approximately $\pm15^\circ$ relative to the nominal viewing direction. A double-pulsed Nd:YAG laser (Quantel EverGreen 532-200) illuminated the spanwise midplane with a light sheet approximately $1~\mathrm{mm}$ thick. Stereo calibration was performed using a dual-plane target, followed by self-calibration to reduce disparity errors.

\begin{figure}
\centering
\includegraphics[width=0.45\textwidth]{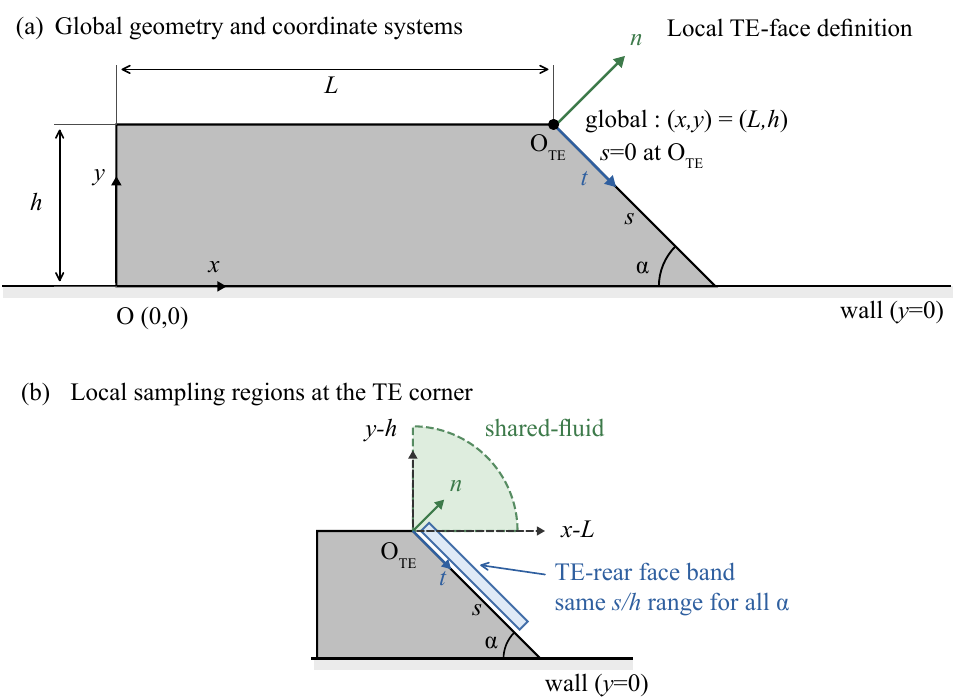}
\caption{Global geometry, coordinate systems, and sampling regions.
(a) Global coordinates $(x,y)$ and the local trailing-edge definition.
The coordinate $s$ denotes distance along the rear face from
$O_{\mathrm{TE}}=(L,h)$, while $t$ and $n$ denote the tangential and
outward-normal directions, respectively.
(b) Shared-fluid region and geometry-following TE-face band used for the
local analysis, with the same $s/h$ range for all $\alpha$.}
\label{coord}
\end{figure}

\begin{figure*}[t]
\centering
\includegraphics[width=0.9\textwidth]{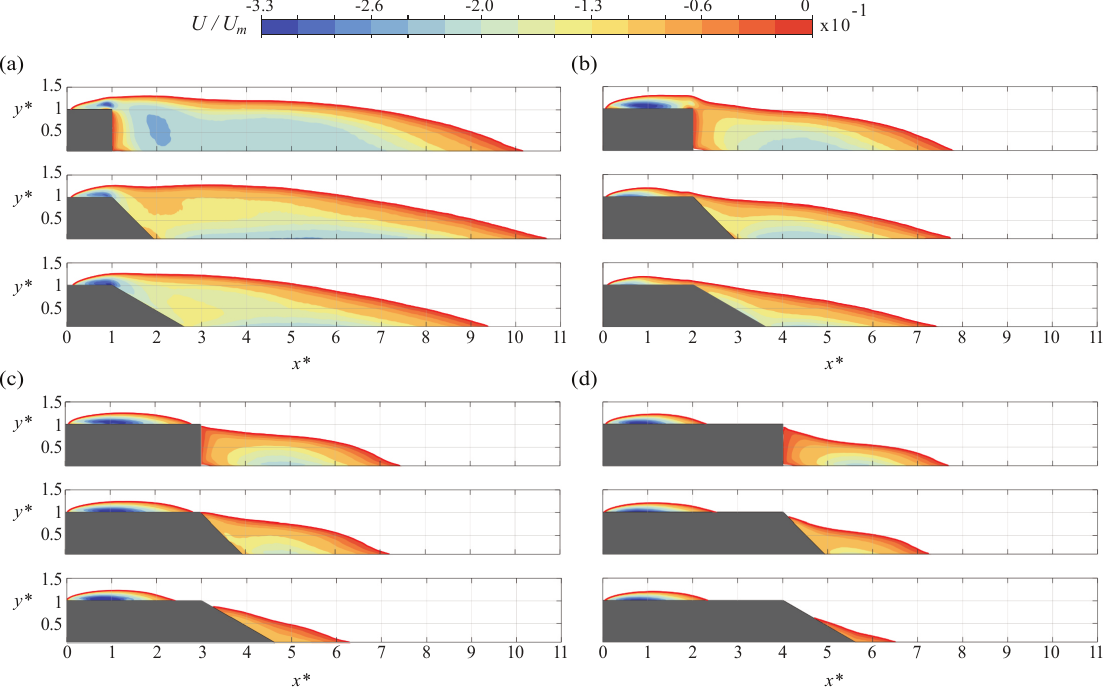}
\caption{Streamwise velocity in the reverse-flow region, $U<0$.
Panels (a)--(d) correspond to $L^*=1$, 2, 3, and 4, respectively. Within each panel, $\alpha=90^\circ$, $45^\circ$, and $30^\circ$ are shown from top to bottom. The red line marks the $U=0$ contour bounding the connected reverse-flow region. The gray area is the model.}
\label{velocity u}
\end{figure*}

Image pairs were acquired at $2~\mathrm{Hz}$ over a field of view of approximately $118~\mathrm{mm}\times54~\mathrm{mm}$ in the streamwise and wall-normal directions, respectively, with an image scale of $46~\mathrm{px\,mm^{-1}}$. The measurement region extended from the foot of the leading edge to approximately $11h$ downstream, covering the upper surface, trailing-edge region and wake for all configurations. Each dataset comprised $4000$--$5000$ image pairs acquired over $33$--$42$ minutes, with intervals between successive snapshots much longer than the convective time $h/\Um$.

We processed the PIV images in DaVis 11.2 (LaVision GmbH) using GPU-based multipass cross-correlation with progressively smaller interrogation  windows and a final window size of $48\times48$ pixels with $75\%$ overlap. The resulting vector spacing was $\Delta=0.2576~\mathrm{mm}$, giving approximately $39$ vectors per protrusion height. The open-source uprime package \cite{jose2026uprime}, was used for subsequent analysis and visualization, including ensemble averaging to obtain mean velocities and Reynolds stresses, and plotting of velocity fields.

A separate PIV evaluation with final interrogation windows of $12\times12$ pixels with $50\%$ overlap yielded a vector spacing of $0.1288~\mathrm{mm}$. This evaluation was used for the mean-flow results in Sec.~\ref{sec:local_te_state}, including the velocity fields and spatially averaged trailing-edge velocities. All other quantitative results, including the Reynolds stresses and turbulent kinetic-energy production, were obtained using the standard processing parameters specified above.

Velocity uncertainties were estimated using the method of \citet{wieneke2015} and were below $1\%$ for all three components, with representative values of approximately $0.5\%$, $0.4\%$ and $0.6\%$ for $U$, $V$ and $W$, respectively. The sensitivity of the spanwise-stress measurements is examined further in \ref{app:spanwise_robustness}.

\subsection{Coordinate Systems and Notation}
\label{sec:coordinates}

Figure~\ref{coord} summarizes the coordinate systems and sampling regions used in the analysis. A Cartesian coordinate system is defined with its origin at the foot of the forward face on the spanwise midplane. The coordinates $x$, $y$ and $z$ denote the streamwise, wall-normal and spanwise directions, respectively, and the stereoscopic PIV measurements are made in the nominal midplane $z=0$. Separation and reattachment locations are reported in this global frame, with $x^*=x/h$ and $y^*=y/h$. For the near-face analysis, a local $(t,n)$ frame defines directions tangential and normal to the rear face, and an arclength coordinate $s$ measures distance along it from the upper trailing-edge corner $(L,h)$ toward the wall. These local coordinates are defined in Sec.~\ref{sec:teface_turbulence}. 

The instantaneous velocity components are denoted $u$, $v$ and $w$, with ensemble means $U=\langle u\rangle$, $V=\langle v\rangle$ and $W=\langle w\rangle$. Angle brackets denote averaging over the instantaneous snapshots acquired for each configuration (Sec.~\ref{sec:piv}), and fluctuations are defined as $u'=u-U$, $v'=v-V$ and $w'=w-W$. In index notation, $u_i$ and $U_i$ denote the instantaneous and mean velocity components, with $i=1,2,3$ corresponding to $x$, $y$ and $z$. The Reynolds stresses are $R_{ij}=\langle u_i'u_j'\rangle$, and the turbulent kinetic energy is $k=R_{ii}/2$, with summation over repeated indices. An overbar denotes a spatial average, with a subscript identifying the sampling region where needed; for example, $\overline{\phi}_{\mathrm{TE}}$ denotes an average over the rear-face sampling band. Lengths are normalized by $h$, velocities by $\Um$, Reynolds stresses and $k$ by $\Um^2$, and turbulent kinetic-energy production by $\Um^3/h$.

\section{Results and Discussion}
\label{sec:results}


\subsection{Reverse-flow regions and reattachment}
\label{sec:global_topology}

Figure~\ref{velocity u} shows the ensemble-mean reverse-flow regions, identified by $U<0$ and bounded by the $U=0$ contour, for all twelve configurations. For $L^*=1$ and $2$, a single connected reverse-flow region extends from the leading edge, over the upper surface and into the downstream wake, before reattaching on the wall. For $L^*=3$ and $4$, the upper-surface separation closes upstream of the trailing edge, and the flow contains two distinct connected reverse-flow regions, one over the upper surface and another downstream of the trailing-edge corner. The wake region is also smaller and has a lower peak reverse velocity than the connected regions observed for the shorter protrusions.

\begin{figure}[!t]
\centering
\includegraphics[width=0.40\textwidth]{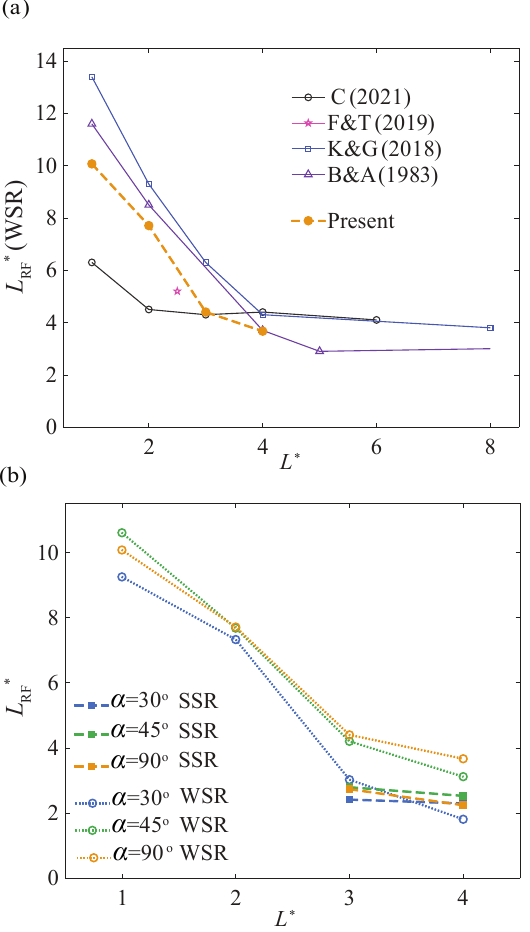}
\caption{Normalized connected reverse-flow extent as a function of protrusion length. (a) Wake-side extent $L_{\mathrm{RF}}^*$ for the present $\alpha=90^\circ$ cases compared with previous rectangular-protrusion measurements. Legend abbreviations: C: Chalmers et al.\
\cite{chalmers2021streamwise}; F\&T: Fang and Tachie~\cite{Fang2019}; K\&G: van der Kindere and Ganapathisubramani~\cite{van2018effect}; B\&A: Bergeles and Athanassiadis~\cite{bergeles1983flow}. (b) Connected reverse-flow extents $L_{\mathrm{RF}}^*$ of the WSR and SSR for the present configurations across $L^*$ and $\alpha$.}
\label{fig:reattachment_comparison}
\end{figure}

We classify the connected reverse-flow regions by the location of their downstream reattachment. A region that reattaches on the upper surface of the protrusion is termed a surface separation region (SSR), whereas one that reattaches on the bottom wall downstream of the protrusion is termed a wake separation region (WSR). The classification refers to the final reattachment location, not to the full spatial extent of the separated region. Table~\ref{tab:reattachment_lengths} uses two quantities to characterize each connected reverse-flow region. The reattachment coordinate, $X_r^*=x_r/h$, gives the streamwise position of the downstream reattachment point in the global frame, measured from the foot of the forward face. The reverse-flow extent, $L_{\mathrm{RF}}^*$, gives the streamwise length of the corresponding connected $U<0$ region, measured from its upstream boundary to the reattachment point.

For $L^*=1$ and $2$, a single connected reverse-flow region extends from the leading edge, over the upper surface, and into the wake before reattaching on the wall. It is therefore classified as a WSR, with no distinct SSR. Because this region begins at the leading edge, $L_{\mathrm{RF}}^*=X_r^*$. For $L^*=3$ and $4$, the reverse flow consists of an upstream SSR that begins at the leading edge and reattaches on the upper surface, and a separate downstream WSR that reattaches on the wall. Thus, $L_{\mathrm{RF}}^*=X_r^*$ for the SSR, whereas the downstream WSR generally has $L_{\mathrm{RF}}^*\neq X_r^*$ because separation does not initiate at the leading edge. For the $45^\circ$ and $90^\circ$ cases, the upstream boundary of the WSR is located at the trailing-edge corner, while for the $30^\circ$ cases it occurs partway down the inclined face, at $x^*\approx3.3$ and $4.7$ for $L^*=3$ and $4$, respectively.

The variation in $L^*$ is clearly more dominant than the variation of the slope angle at a fixed length. For $L^*=3$ and $4$, the SSR reattaches at $X_r^*\approx2.3$--$2.8$ for all three angles, suggesting that once sufficient upper-surface length is available, its reattachment is governed primarily by the leading-edge separation. This range lies beyond the trailing edge for $L^*=2$, consistent with the absence of a distinct SSR for the shorter protrusions.

The rear-face angle mainly affects the extent of the WSR, particularly for the longer protrusions. At $L^*=4$, the WSR extends over $1.8h$ for $\alpha=30^\circ$ and $3.7h$ for $\alpha=90^\circ$. Part of this difference arises because the WSR begins farther downstream along the inclined face in the $30^\circ$ case, so the change in its extent is larger than the shift in its wall reattachment position. Smaller differences between angles require caution. For the WSR at $L^*=2$ and the SSR at $L^*=3$--$4$, these differences are comparable to the sensitivity of the estimated reattachment position, which reaches $0.40h$ and $0.31h$, respectively (see ~\ref{app:reattachment_sensitivity}). These small differences therefore do not establish a clear effect of $\alpha$. Overall, $L^*$ primarily determines whether the flow contains a single connected reverse-flow region or separate SSR and WSR regions, while $\alpha$ mainly modifies the shape and extent of the WSR.

Figure~\ref{fig:reattachment_comparison}(a) compares the present $L_{\mathrm{RF}}^*$ values with previous measurements reported in the literature for rectangular protrusions. For the $\alpha=90^\circ$ cases, $L_{\mathrm{RF}}^*=10.2$, $7.8$, $4.4$, and $3.7$ for $L^*=1$, 2, 3, and 4, respectively. We included results from \citet{bergeles1983flow, van2018effect, Fang2019, chalmers2021streamwise} using the same normalization. Minor differences between the present results and the literature may also reflect differences in Reynolds number and incoming boundary-layer conditions, which are constant within the present dataset. Figure~\ref{fig:reattachment_comparison}(b) shows the WSR and SSR extents for the present cases. 

\begin{table}[t]
\centering
\begin{tabular}{c c c c c c}
\toprule
& & \multicolumn{2}{c}{WSR} & \multicolumn{2}{c}{SSR} \\
\cmidrule(lr){3-4}\cmidrule(lr){5-6}
$L^*$ & $\alpha$ ($^\circ$) &
$L_{\mathrm{RF}}^*$ & $X_r^*$ &
$L_{\mathrm{RF}}^*$ & $X_r^*$ \\
\midrule
\midrule
\multirow{3}{*}{1}
 & 30 & 9.4 & 9.4 & -- & -- \\
 & 45 & 10.7 & 10.7 & -- & -- \\
 & 90 & 10.2 & 10.2 & -- & -- \\
\midrule
\multirow{3}{*}{2}
 & 30 & 7.4 & 7.4 & -- & -- \\
 & 45 & 7.7 & 7.7 & -- & -- \\
 & 90 & 7.8 & 7.8 & -- & -- \\
\midrule
\multirow{3}{*}{3}
 & 30 & 3.0 & 6.3 & 2.4 & 2.4 \\
 & 45 & 4.2 & 7.2 & 2.8 & 2.8 \\
 & 90 & 4.4 & 7.4 & 2.8 & 2.8 \\
\midrule
\multirow{3}{*}{4}
 & 30 & 1.8 & 6.5 & 2.3 & 2.3 \\
 & 45 & 3.1 & 7.2 & 2.5 & 2.5 \\
 & 90 & 3.7 & 7.7 & 2.3 & 2.3 \\
\bottomrule
\end{tabular}
\caption{Connected reverse-flow extents and reattachment coordinates. $L_{\mathrm{RF}}^*$ is the streamwise extent of an individual connected reverse-flow (RF) region, measured from its own upstream boundary to its downstream reattachment point. $X_r^*=x_r/h$ is the corresponding global reattachment coordinate measured from the foot of the forward face. WSR denotes a region that reattaches on the wall ($y=0$), while SSR denotes
a region that reattaches on the upper surface ($y=h$). Only a WSR is present for $L^*=1$ and $2$, while separate SSR and WSR regions appear for $L^*=3$ and $4$.}
\label{tab:reattachment_lengths}
\end{table}

\subsection{Mean flow at the trailing edge}
\label{sec:local_te_state}
\begin{figure*}[t]
\centering
\includegraphics[width=0.85\textwidth]{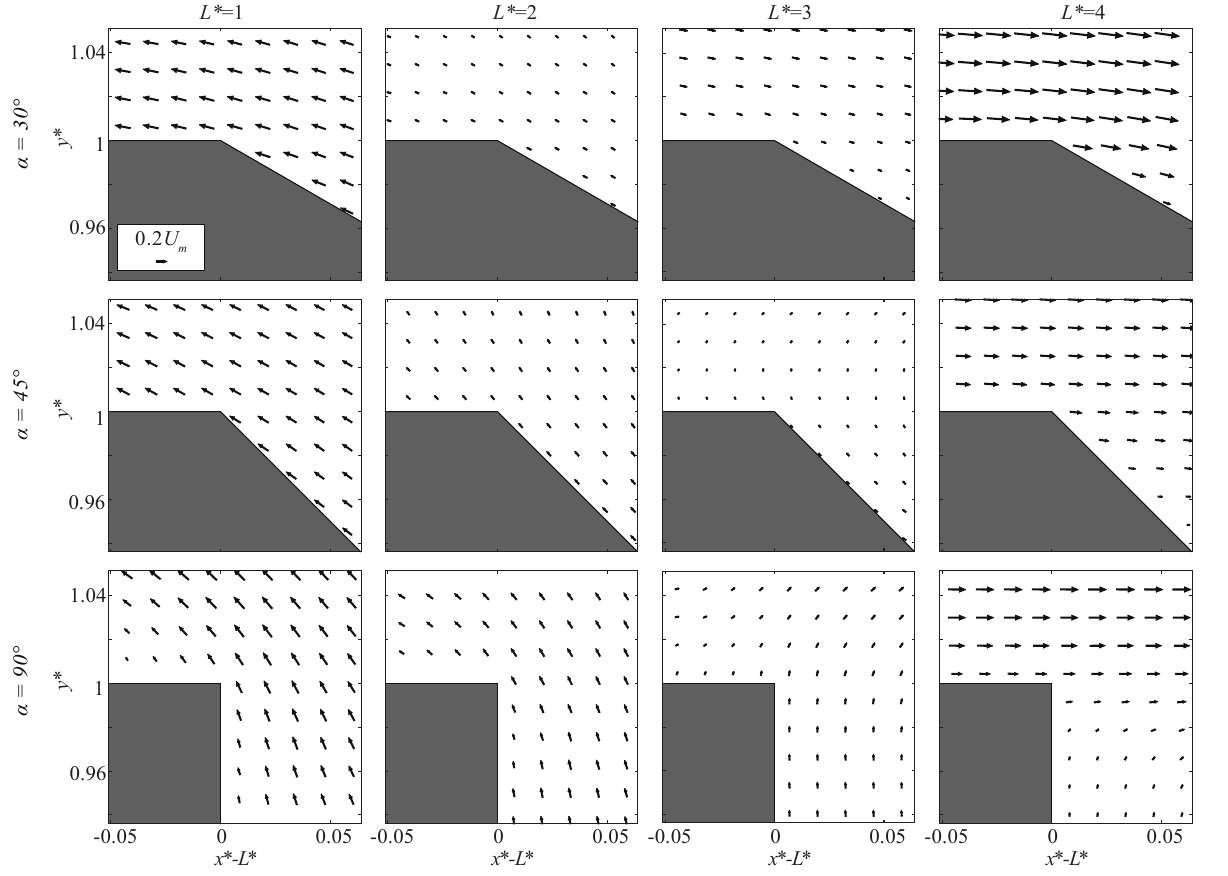}
\caption{Time-averaged velocity field around the trailing edge for all twelve configurations. Columns correspond to model length $L^*$, and rows to trailing-edge angle $\alpha$. Each panel shows the region around the trailing-edge corner covered by the averaging window used to compute the spatially averaged velocity vector (Fig.~\ref{polar_vector}). Arrows show the normalized in-plane velocity at each grid point, drawn to a common scale in all panels; the reference arrow corresponds to $0.2\,U_m$. The gray region denotes the model.
}
\label{velocity TE}
\end{figure*}

Figure~\ref{velocity TE} shows how protrusion length changes the mean flow crossing the trailing-edge corner. For $L^*=1$, strong return flow rises along the rear face and continues upstream over the upper surface. As $\alpha$ increases, this flow leaves the corner at a steeper upward angle, departing farther from the upper surface. The return flow is substantially weaker at $L^*=2$. For $L^*=3$, reattachment occurs shortly upstream of the trailing edge, leaving little distance for the forward flow to recover before reaching the corner. The local velocity therefore remains small, and its direction depends strongly on the rear-face angle. The flow turns downward for $\alpha=30^\circ$, whereas upward motion dominates for $\alpha=45^\circ$ and $90^\circ$. By $L^*=4$, the greater distance between reattachment and the trailing edge allows stronger, predominantly streamwise flow to develop. The shallow face deflects this flow slightly downward. At the vertical face, the flow remains nearly horizontal above weaker upward motion below the corner. Thus, the effect of rear-face inclination depends on whether the corner lies within the return flow, close to upper-surface reattachment, or farther downstream where the forward flow has recovered.

\begin{figure}[h!]
\centering
\includegraphics[width=0.40\textwidth]{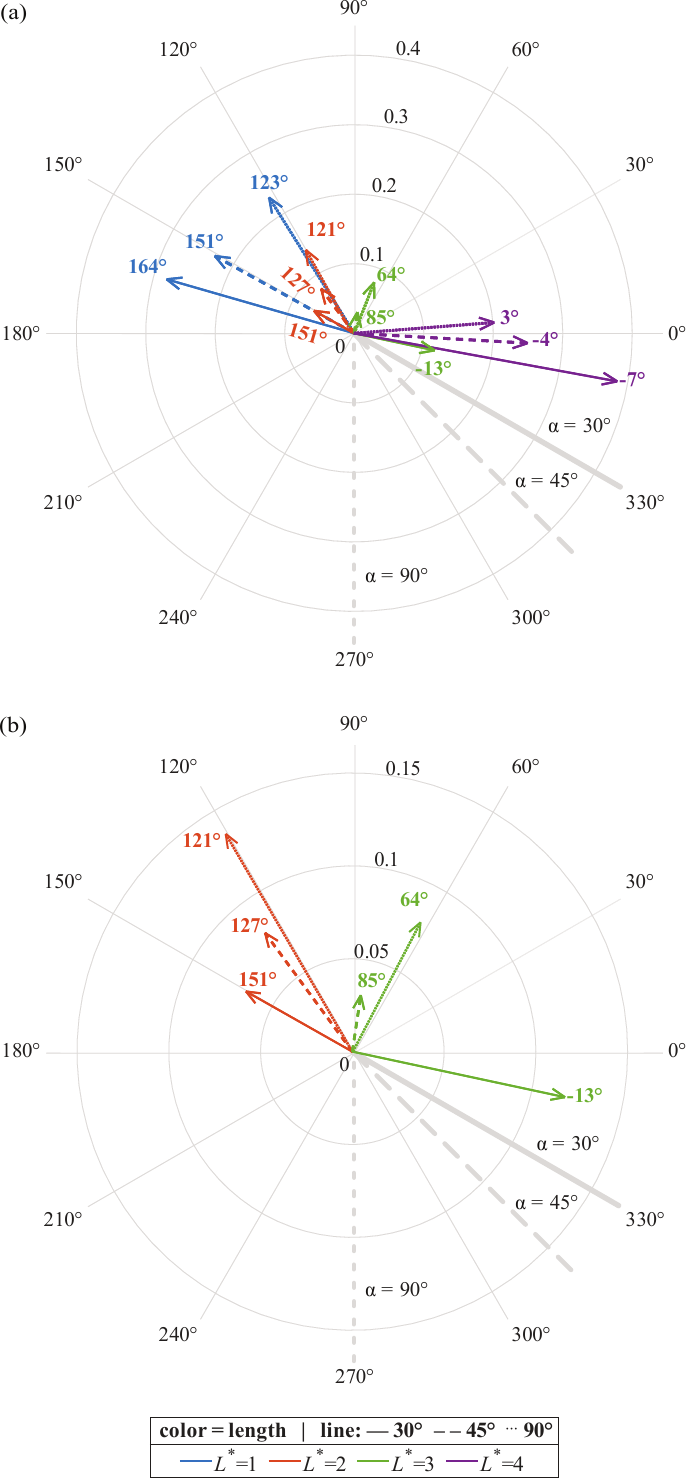}
\caption{Area-averaged mean velocity vector in the neighborhood of the TE corner. (a) Comparison of lengths $L^*=1-4$ and (b) zoomed-in comparison of $L^*=2-3$. Radial position denotes $\left|\overline{\mathbf{U}}_{2D,A}\right|/\Um$ and polar angle denotes $\theta_A$. Color denotes $L^*$ and line style denotes the TE slope $\alpha$ (solid $30^\circ$, dashed $45^\circ$, dotted $90^\circ$). The value at each arrow tip is $\theta_A$ in degrees.}
\label{polar_vector}
\end{figure}

Figure~\ref{polar_vector} quantifies the local mean flow by averaging the velocity components over a square window with a side length of $\sim 0.13h$, centered at the TE corner. The window contains a $10\times10$ grid of vectors from the finer reconstructed velocity fields ($\Delta=0.1288mm$). The spatial average of each ensemble-mean velocity component over this window is calculated as
\begin{equation}
\overline{\phi}_A= \frac{1}{N_{\mathrm{valid}}}\sum_{p\in\mathcal V_A}\phi_p,
\qquad
\phi\in\{U,V\},
\label{eq:te_area_average}
\end{equation}
where $\mathcal V_A$ is the set of valid, unmasked grid points within the window, $N_{\mathrm{valid}}$ is their number, and $\phi_p$ is the velocity component at grid point $p$. The magnitude and direction of the spatially averaged velocity vector are
\begin{equation}
\begin{aligned}
\left|\overline{\mathbf{U}}_{2D,A}\right|&=\left(\overline{U}_A^{\,2}+\overline{V}_A^{\,2}\right)^{1/2},\\
\theta_A&=\operatorname{atan2}\left(\overline{V}_A,\overline{U}_A\right).
\end{aligned}
\label{eq:te_mean_polar}
\end{equation}
Here, $\operatorname{atan2}$ denotes the two-argument arctangent, which accounts for the signs of both velocity components to determine the flow direction. The angle $\theta_A$ is measured counterclockwise from the downstream direction and reported in degrees. In Fig.~\ref{polar_vector}, the radial coordinate gives $\left|\overline{\mathbf{U}}_{2D,A}\right|/\Um$, and the polar angle gives $\theta_A$. Panel~(b) enlarges the weaker vectors for $L^*=2$ and $3$.

For $L^*=1$, the vectors show strong return flow, with magnitudes of $0.23$--$0.28\,\Um$ and directions between $123^\circ$ and $164^\circ$. The shallowest face produces the strongest and most nearly streamwise return flow. The upslope directions of the rear faces are $150^\circ$, $135^\circ$, and $90^\circ$ for $\alpha=30^\circ$, $45^\circ$, and $90^\circ$, respectively. At all three angles, the vector is rotated by $14^\circ$--$33^\circ$ from the upslope direction toward the upstream horizontal ($180^\circ$), consistent with the return flow turning to follow the upper surface as it crosses the corner.

At $L^*=2$, the flow remains reversed, but the vector magnitudes decrease to approximately $0.07$--$0.13\,\Um$. The directions are $151^\circ$, $127^\circ$, and $121^\circ$ for $\alpha=30^\circ$, $45^\circ$, and $90^\circ$, respectively. These lie within $31^\circ$ of the corresponding upslope directions, $150^\circ$, $135^\circ$, and $90^\circ$, indicating closer alignment with the rear face than at $L^*=1$. The magnitude ordering is also reversed, with the strongest return flow at the vertical face and the weakest at the shallowest face.

At $L^*=3$, the direction varies most strongly with $\alpha$, although the local mean velocity remains weak. The velocity vector for the $30^\circ$ case has a magnitude of approximately $0.12\,\Um$ and points slightly downward at $-13^\circ$. The vectors for the $45^\circ$ and $90^\circ$ cases have smaller magnitudes, approximately $0.03$ and $0.08\,\Um$, and point predominantly upward at $85^\circ$ and $64^\circ$, respectively. These differences occur close to upper-surface reattachment, where the forward flow has had little distance to recover before reaching the corner.

By $L^*=4$, all three vectors show stronger forward flow, with magnitudes of approximately $0.2$--$0.4\,\Um$ that increase as the rear face becomes shallower. Their directions occupy a much narrower range, from $-7^\circ$ to $3^\circ$. Thus, changing $\alpha$ produces an approximately twofold difference in velocity magnitude but only a $10^\circ$ change in direction. The flow crosses the corner predominantly in the streamwise direction, with modest downward deflection toward the inclined faces and a slight upward deflection above the vertical face.

\subsection{Turbulence at the trailing-edge face}
\label{sec:teface_turbulence}

The turbulence close to the rear face is sampled using the geometry-following band shown in Fig.~\ref{coord}(b). Position along the face is described by $s$, with $s=0$ at the upper TE corner and increasing toward the wall, while $t$ and $n$ denote the corresponding tangential and outward-normal directions. The normalized coordinate is defined as $s^*=s/h$. The baseline sampling band is three vectors wide, extends from $2\Delta$ to $4\Delta$ normal to the surface, and spans $0.05\leq s^*\leq0.90$. For a given $L^*$, the same range of $s^*$ is used for all three rear-face angles, so the comparison is made at equal distance from the TE corner along each face. Changing $\alpha$ therefore rotates the sampling location about $O_{\mathrm{TE}}$ while preserving its face-relative position. The $\alpha=90^\circ$ face is the shortest, with length $h$, and thus limits the
common range. The comparison is therefore based on equal distance from the TE corner rather than on an equal fraction of the full rear-face length.

The flow sampled by this band depends on the local TE flow shown in Fig.~\ref{polar_vector}, which is reversed for $L^*=1$ and $2$, transitional at $L^*=3$, and predominantly forward at $L^*=4$. Accordingly, for $L^*=1$ and $2$, the TE-face band primarily samples the near-face recirculating flow within the WSR. For $L^*=3$ and $4$, the SSR and WSR are distinct and the band samples the near-face flow associated with the WSR. In the $30^\circ$ cases, separation occurs partway down the inclined face, so part of the band lies within the near-surface forward-flow region before entering the WSR. The shared-fluid comparison in Sec.~\ref{sec:geometric_control} complements this analysis by comparing fixed global locations. Sensitivity to the band position is reported in ~\ref{app:sampling_robustness}.

To describe the turbulence relative to the rear face, the in-plane Reynolds-stress tensor is rotated into the face-attached $(t,n)$ basis (Fig.~\ref{coord}), where $t$ and $n$ denote the tangential and outward-normal directions, respectively,
\begin{equation}
\mathbf{R}^{(t,n)}=\mathbf{Q}\mathbf{R}_{2D}\mathbf{Q}^{\top},
\qquad
\mathbf{Q}(\alpha)=
\begin{bmatrix}
\cos\alpha & -\sin\alpha\\
\sin\alpha & \cos\alpha
\end{bmatrix}.
\end{equation}
The components $R_{tt}$, $R_{nn}$, and $R_{zz}$ represent the face-tangential, face-normal, and spanwise fluctuation energies, respectively. Their ratios to $2k$ quantify the directional partition of turbulent kinetic energy, with $1/3$ in each direction corresponding to isotropy.

\begin{figure*}[h]
\centering
\includegraphics[width=0.90\textwidth]{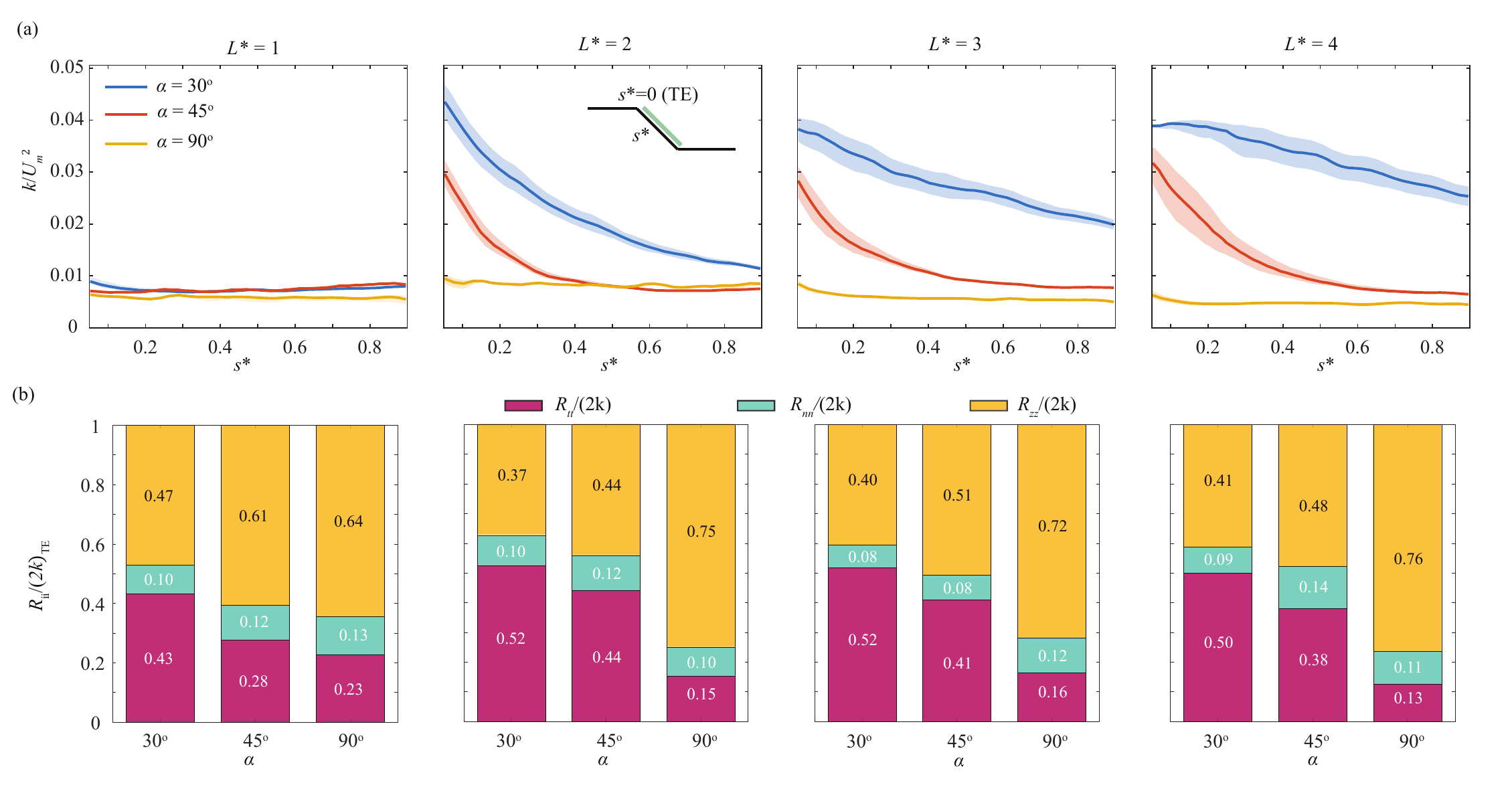}
\caption{TE-face turbulence magnitude and directional energy partition for $L^*=1$--4. (a) Turbulent kinetic energy, $k/\Um^2$, along the TE-face coordinate $s^*=s/h$. Shaded bands indicate the variation obtained when the normal sampling location is shifted between $2\Delta$ and $4\Delta$. (b) TE-face-band-averaged partition of turbulent energy into the face-tangential, face-normal, and spanwise components, $\overline{R_{tt}/(2k)}_{\mathrm{TE}}$,  $\overline{R_{nn}/(2k)}_{\mathrm{TE}}$, and $\overline{R_{zz}/(2k)}_{\mathrm{TE}}$.}
\label{stereo_turbulence_TEface}
\end{figure*}

Figure~\ref{stereo_turbulence_TEface}(a) shows the variation of turbulent kinetic energy along the TE-face band. For $L^*=1$, turbulent kinetic energy is relatively low for all three angles and depends only weakly on $\alpha$. From $L^*=2$ onward, a clear ordering emerges: the $30^\circ$ cases have the largest $k/U_m^2$, the $45^\circ$ cases are intermediate, and the $90^\circ$ cases remain consistently lower. The $30^\circ$ and $45^\circ$ profiles also decrease with increasing $s^*$, whereas the $90^\circ$ profiles are comparatively weak and nearly uniform. Thus, once the local flow at the TE reorganizes, rear-face angle has a strong influence on the magnitude of the near-face turbulent fluctuations. This trend is consistent with the changes in the local TE flow described in Sec.~\ref{sec:local_te_state}. The spatial production fields in Sec.~\ref{sec:spatial_production} show where this turbulent energy is generated locally.

Figure~\ref{stereo_turbulence_TEface}(b) shows the corresponding directional partition of turbulent kinetic energy. The face-normal contribution remains small in all cases, with $R_{nn}/(2k)\approx0.08$--$0.14$. The main change with $\alpha$ is a redistribution between the tangential and spanwise components. At $30^\circ$, the tangential and spanwise contributions are comparable, with $R_{tt}/(2k)\approx0.43$--$0.52$. As $\alpha$ increases, the tangential fraction decreases while the spanwise fraction increases. For $90^\circ$, $R_{zz}/(2k)$ reaches $0.64$--$0.76$, while $R_{tt}/(2k)$ falls to $0.13$--$0.23$. This partition varies much more systematically with $\alpha$ than with $L^*$.

Together, the two panels show that turbulence magnitude and directional partition respond differently to the rear-face geometry. The $30^\circ$ cases generally have the largest total turbulent kinetic energy, whereas the $90^\circ$ cases have the smallest $k$. At every $L^*$, however, the spanwise component carries the largest share of $k$ at the vertical face, with $\overline{R_{zz}/(2k)}_{\mathrm{TE}} = 0.64$--$0.76$, about twice the isotropic value and the highest of the three angles. This dominance holds only in relative terms, since at $L^*=1$--$2$, the absolute spanwise stress $\overline{R_{zz}}_{\mathrm{TE}}$ at $90^\circ$ is comparable to that at the inclined faces. For $L^*=3$--$4$, where $k$ at $90^\circ$ is about 5--7 times smaller than at $30^\circ$, $\overline{R_{zz}}_{\mathrm{TE}}$ is about 2.5--4 times smaller. The robustness of the large spanwise fraction at $90^\circ$ is examined in ~\ref{app:spanwise_robustness}. The spanwise contribution also enters the divergence-associated production through Eq.~\eqref{eq:Pdiv_fz}.

\subsection{Turbulent kinetic-energy production at the trailing-edge face}
\label{sec:production}

\subsubsection{Midplane form of the production}
\label{sec:midplane}

The turbulent kinetic-energy production is given by the  neagative contraction of the Reynolds-stress tensor with the mean velocity-gradient tensor,
\begin{equation}
P = -\langle u_i'u_j'\rangle \frac{\partial U_i}{\partial x_j}.
\end{equation}
The stereoscopic PIV resolves all three velocity components, all six Reynolds stresses, and the in-plane velocity gradients within the measurement plane. Since the measurement is planar, derivatives in the spanwise direction are not measured directly.

The production at the nominal midplane is evaluated assuming statistical reflection symmetry about $z=0$. Under this symmetry, $U$ and $V$ are even functions of $z$, while $W$ is odd \citep{pope2000turbulent}. Likewise, correlations containing exactly one spanwise fluctuation, $\langle u'w'\rangle$ and $\langle v'w'\rangle$, are odd in $z$, whereas $\langle w'w'\rangle$ is even. Since odd functions, including the $z$-derivatives of even functions, vanish at $z=0$, the midplane satisfies 

\begin{equation}
\frac{\partial U}{\partial z} \simeq \frac{\partial V}{\partial z} \simeq 0,
\qquad
\langle u'w'\rangle \simeq \langle v'w'\rangle \simeq 0.
\label{eq:midplane_symmetry}
\end{equation}

The measured departures of $\langle u'w'\rangle$ and $\langle v'w'\rangle$ from zero are used to assess this assumption in
Appendix~A.2. The remaining spanwise gradient at the midplane, $\partial W/\partial z$, is not measured directly but follows from mean continuity, $\partial U/\partial x + \partial V/\partial y + \partial W/\partial z = 0$, which holds exactly for incompressible flow. Defining the measured in-plane
divergence

\begin{equation}
D_{2D} = \frac{\partial U}{\partial x} + \frac{\partial V}{\partial y},
\label{eq:D2D}
\end{equation}
gives
\begin{equation}
\frac{\partial W}{\partial z} = -D_{2D}.
\label{eq:continuity}
\end{equation}

The reconstructed midplane production can then be separated into an in-plane contribution,

\begin{equation}
P_{2D} = -\mathbf{R}_{2D}:\mathbf{S}_{2D},
\label{eq:P2D}
\end{equation}

and a spanwise contribution,

\begin{equation}
\Pzz = -R_{zz}\frac{\partial W}{\partial z} = R_{zz}\Dtd.
\end{equation}

Here $\mathbf{S}_{2D}$ is the symmetric in-plane mean strain-rate tensor, with

\begin{equation}
S_{xx}=\frac{\partial U}{\partial x},
\qquad
S_{yy}=\frac{\partial V}{\partial y},
\qquad
S_{xy}=S_{yx} = \frac{1}{2} \left( \frac{\partial U}{\partial y} + \frac{\partial V}{\partial x} \right).
\end{equation}

The spanwise stress $R_{zz}$ is measured directly by stereoscopic PIV, whereas $\partial W/\partial z$ is obtained from continuity.

The measurable departures from the symmetry conditions in Eq.~\eqref{eq:midplane_symmetry} are assessed in ~\ref{app:spanwise_robustness}. Over the baseline TE-face band, their contribution ranges from $0.4\%$ to $10\%$ of the reconstructed production, with the largest relative values occurring at $L^*=1$, where the net production is small. The remaining spanwise derivatives are evaluated under the midplane symmetry assumption.

\subsubsection{Invariant decomposition of the in-plane production}
\label{sec:invariant}

The in-plane production $P_{2D}$ in Eq.~\eqref{eq:P2D} is the double contraction of two symmetric second-order tensors and is invariant under coordinate rotation \citep{pope2000turbulent}. The face-attached $(t,n)$ basis introduced in Sec.~\ref{sec:teface_turbulence} can therefore be used without changing $P_{2D}$. The strain-rate tensor is transformed using the same rotation as the Reynolds-stress tensor,

\begin{equation}
\mathbf{S}_{tn} = \mathbf{Q}\mathbf{S}_{2D}\mathbf{Q}^{\top}.
\end{equation}

We split each tensor into an isotropic part and a traceless deviatoric part,

\begin{equation}
\mathbf{R}_{2D} = C_R\mathbf{I} + \mathbf{R}', \qquad
\mathbf{S}_{2D} = C_S\mathbf{I} + \mathbf{S}',
\end{equation} 

Here $\mathbf{I}$ is the $2\times2$ identity tensor. The primes denote the traceless deviatoric parts $\mathbf{R}'$ and $\mathbf{S}'$, obtained by subtracting the isotropic parts $C_R\mathbf{I}$ and $C_S\mathbf{I}$. The isotropic coefficients are the half-traces

\begin{equation}
C_R = \frac{R_{tt}+R_{nn}}{2},
\qquad
C_S = \frac{S_{tt}+S_{nn}}{2}.
\end{equation}

The deviatoric magnitudes are

\begin{equation}
\begin{aligned}
r_R &= \sqrt{ \left( \frac{R_{tt}-R_{nn}}{2} \right)^2 + R_{tn}^2 },\\
r_S &= \sqrt{ \left( \frac{S_{tt}-S_{nn}}{2} \right)^2 + S_{tn}^2 },
\end{aligned}
\end{equation}

and the corresponding principal directions are

\begin{equation}
\begin{aligned}
\theta_R &= \tfrac{1}{2} \operatorname{atan2} \left( 2R_{tn}, R_{tt}-R_{nn} \right),\\
\theta_S &= \tfrac{1}{2} \operatorname{atan2} \left( 2S_{tn}, S_{tt}-S_{nn} \right). 
\end{aligned}
\end{equation}

The angles $\theta_R$ and $\theta_S$ are measured from the face-tangential direction $t$, positive toward the outward normal $n$. These quantities give the usual spectral representation of a two-dimensional symmetric tensor, equivalent to its Mohr-circle representation. Principal-axis descriptions are widely used to examine the relation between Reynolds stress and mean strain in turbulent flows \citep{pope1975effective,schmitt2007boussinesq,lengani2020principal}.

Because a principal axis is an unoriented line, the local stress--strain misalignment is defined as

\begin{equation}
\Delta\theta = \cos^{-1} \left[ \left|\cos(\theta_R-\theta_S)\right|\right],
\qquad
0\leq\Delta\theta\leq90^\circ.
\end{equation}

For the deviatoric parts,

\begin{equation}
\mathbf{R}':\mathbf{S}' = 2r_Rr_S\cos(2\Delta\theta),
\end{equation}

The in-plane production therefore becomes

\begin{equation}
\begin{aligned}
P_{2D} &= \underbrace{-2C_RC_S}_{\Ptrace} +\underbrace{\left[-2r_Rr_S\cos(2\Delta\theta)\right]}_{\Palign},\\
\Palign &=-2r_Rr_S\cos(2\Delta\theta).\end{aligned}\label{eq:P_trace_align}
\end{equation}

Thus,

\begin{equation}
P_{2D}=\Ptrace+\Palign.
\end{equation}

Equation~\eqref{eq:P_trace_align} decomposes the original tensor contraction exactly, preserves its rotational invariance, and introduces no additional modeling assumption. The trace term describes the interaction between the isotropic parts of the in-plane stress and strain tensors, while the alignment term describes the interaction between their deviatoric parts.

The alignment contribution separates into amplitude and orientation factors. The product $r_Rr_S$ is non-negative and measures the local stress--strain amplitude, while $\cos(2\Delta\theta)$ depends only on the relative principal-axis orientation and determines the sign of $\Palign$.
Consequently,

\begin{equation}
\Palign<0 \quad\mathrm{for}\quad
\Delta\theta<45^\circ,\qquad
\Palign>0\quad\mathrm{for}\quad
\Delta\theta>45^\circ.
\end{equation}

The limiting orientation connects this decomposition to linear eddy-viscosity modeling. For a positive eddy viscosity, the Boussinesq relation gives

\begin{equation}
\mathbf{R}' = -2\nu_t\mathbf{S}'
\end{equation} 
\citep{pope1975effective,schmitt2007boussinesq}.

The tensors then share the same principal lines but have oppositely ordered eigenvalues. Under the present convention, which identifies the directions associated with the larger eigenvalues, the direction of maximum Reynolds stress is therefore perpendicular to the direction of maximum extensional strain, giving $\Delta\theta=90^\circ$ and positive alignment production. The difference $90^\circ-\Delta\theta$ therefore quantifies the departure from the principal-axis relation implied by a positive linear eddy-viscosity model \citep{lengani2020principal,corsini2025transport}.

For the TE-face average, the amplitude and orientation factors separate as

\begin{equation}
\begin{aligned}
\overline{\Palign}_{\mathrm{TE}} &= -\ARS \overline{\cos(2\Delta\theta)}_{RS},\\
\ARS &= 2\overline{r_Rr_S}_{\mathrm{TE}},
\end{aligned}
\label{eq:Palign_factorized}
\end{equation}

where

\begin{equation}
\overline{\cos(2\Delta\theta)}_{RS} = \frac{ \overline{r_Rr_S\cos(2\Delta\theta)}_{\mathrm{TE}}}{\overline{r_Rr_S}_{\mathrm{TE}}}
\end{equation}

is the orientation factor weighted by the local stress--strain amplitude. The subscript $RS$ denotes weighting by $r_Rr_S$. The factorization is exact when both factors are evaluated over the same sampling region. The orientation factor determines the sign of the band-averaged alignment contribution, while $\ARS$ gives the available amplitude.

\subsubsection{Alignment and divergence-associated contributions}
\label{sec:align_div}

The trace and spanwise contributions are linked through incompressibility. Since

\begin{equation}
C_S = \frac{S_{xx}+S_{yy}}{2}=\frac{\Dtd}{2},
\end{equation}

and the in-plane trace of the Reynolds-stress tensor is invariant under
rotation,

\begin{equation}
\Ptrace=-\frac{R_{xx}+R_{yy}}{2}\Dtd.
\end{equation}

At the symmetry plane, $\partial W/\partial z=-\Dtd$, and therefore

\begin{equation}
\Pzz=-R_{zz}\frac{\partial W}{\partial z}=R_{zz}\Dtd.
\end{equation}

These relations follow directly from the Reynolds-stress production term and incompressible continuity. In strictly two-dimensional incompressible flow, $\Dtd=0$ and both contributions vanish. Because $\Ptrace$ and $\Pzz$ contain the same measured in-plane divergence, they can be grouped as

\begin{equation}
\Pdiv \equiv \Ptrace+\Pzz = \left[ R_{zz}-\frac{R_{xx}+R_{yy}}{2}\right]\Dtd.
\end{equation}

Using

\begin{equation}
2k =R_{xx}+R_{yy}+R_{zz},\qquad
f_z=\frac{R_{zz}}{2k},
\end{equation}

gives

\begin{equation}
\Pdiv =k(3f_z-1)\Dtd.
\label{eq:Pdiv_fz}
\end{equation}

The label ``divergence-associated'' indicates that both terms are proportional to the measured in-plane divergence $D_{2D}$. It does not denote a separate physical mechanism. The contribution $P_{\text{div}}$ follows directly from the production term and continuity, without an additional modeling assumption.
The reconstructed midplane production is therefore

\begin{equation}
P = \Palign+\Pdiv.
\end{equation}

The two contributions serve different purposes in the following analysis. $\Palign$ represents the in-plane deviatoric stress--strain interaction and is used as the primary diagnostic of the local production mechanism. $\Pdiv$ contains the remaining trace and spanwise contributions required to complete the reconstructed midplane production.

For incompressible flow, a nonzero $\Dtd$ can represent genuine spanwise mean stretching at the midplane. However, $\Dtd$ is obtained from the residual sum of two measured spatial derivatives and can therefore be particularly sensitive to differentiation uncertainty and cancellation between the in-plane normal gradients \citep{sciacchitano2019}.

Equation~\eqref{eq:Pdiv_fz} also connects $\Pdiv$ to the spanwise energy partition shown in Fig.~\ref{stereo_turbulence_TEface}(b). The factor $(3f_z-1)$ is largest for the vertical face because a larger fraction of its turbulent kinetic energy is carried by the spanwise component. This factor alone does not determine the magnitude or sign of $\Pdiv$, which also depends on $k$ and $\Dtd$. Sensitivity to spanwise-stress uncertainty, residual divergence, and finite-span effects is examined in ~\ref{app:spanwise_robustness}. Accordingly, $\Palign$ is used below as the primary diagnostic of the production mechanism. $\Pdiv$ represents production by mean spanwise stretching or compression at the midplane, $\partial W/\partial z=-D_{2D}$, acting on the difference between the spanwise and in-plane normal stresses. Because $\partial W/\partial z$ is inferred from continuity through the in-plane divergence rather than measured, $\Pdiv$ is reported to complete the midplane production budget but is not used to interpret the mechanism.
\begin{figure*}[!t]
\centering
\includegraphics[width=0.9\textwidth]{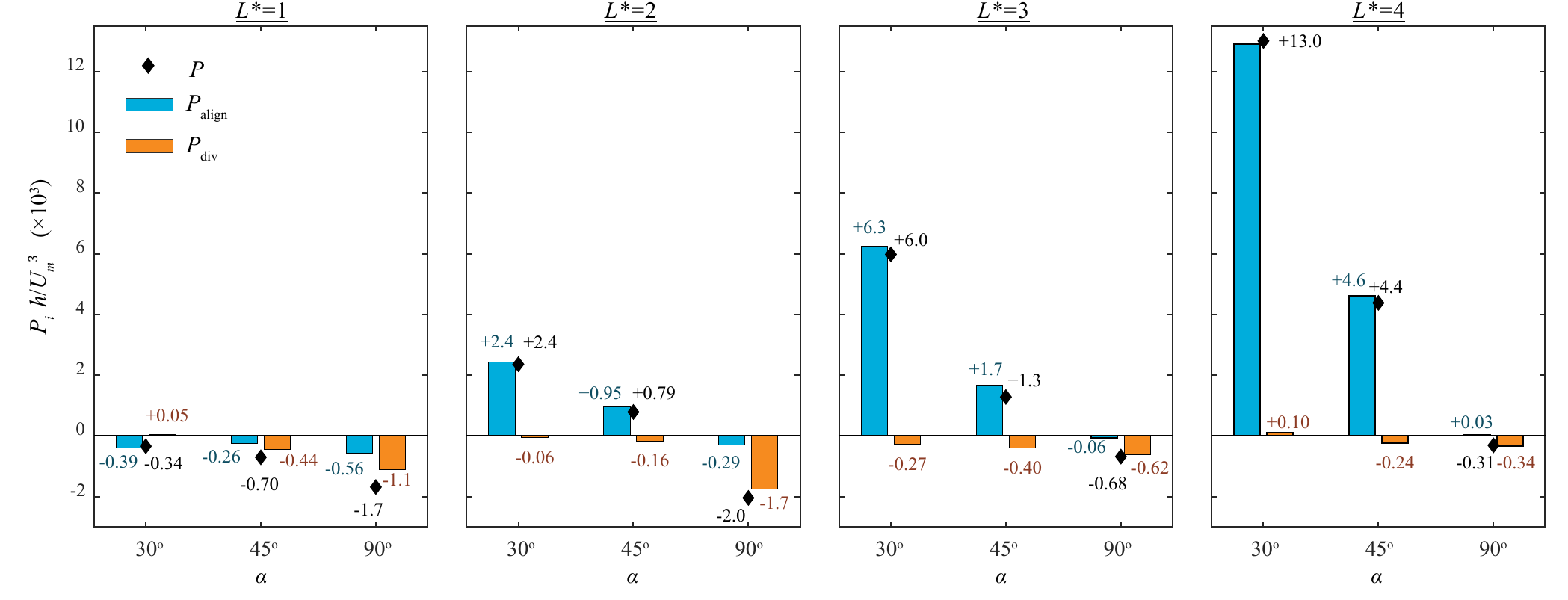}
\caption{TE-face turbulent kinetic-energy production for $L^*=1$--4, from left to right. Bars show the stress--strain alignment contribution $\overline{\Palign}_{\mathrm{TE}}$ and the divergence-associated contribution $\overline{\Pdiv}_{\mathrm{TE}}$. Black diamonds show the reconstructed midplane total $\overline{P}_{\mathrm{TE}}=\overline{\Palign}_{\mathrm{TE}}+\overline{\Pdiv}_{\mathrm{TE}}$. Values are TE-face band averages normalized by $\Um^3/h$ and multiplied by $10^3$.}
\label{Production_decomposition}
\end{figure*}

\begin{table*}[!b]
\centering
\caption{Principal-axis statistics over the matched TE-face band. All weighted quantities use the local stress--strain amplitude $r_Rr_S$, denoted by the subscript $RS$. The mean orientations $\overline{\theta_R}_{RS}$ and $\overline{\theta_S}_{RS}$ are axial means calculated by doubling the $180^\circ$-periodic angles, averaging them with weights $r_Rr_S$, and then halving the resulting angle. $\overline{\Delta\theta}_{RS}$ and $\widetilde{\Delta\theta}$ are the weighted mean and median of the local stress--strain misalignment, respectively, and $F_{>45}$ is the weighted fraction for which $\Delta\theta>45^\circ$. Since $\overline{\Delta\theta}_{RS}$ is obtained from the local misalignment before averaging, it is not generally equal to the difference between the two mean orientations. The final two columns separate the orientation factor $\overline{\cos(2\Delta\theta)}_{RS}$ from the stress--strain amplitude $\ARS=2\overline{r_Rr_S}_{\mathrm{TE}}$.}
\label{tab:angle_summary}
\small
\begin{tabular}{ccrrrrrrr}
\toprule
$L^*$ &
$\alpha$ ($^\circ$) &
$\overline{\theta_R}_{RS}$ &
$\overline{\theta_S}_{RS}$ &
$\overline{\Delta\theta}_{RS}$ &
$\widetilde{\Delta\theta}$ &
$F_{>45}$ (\%) &
$\overline{\cos 2\Delta\theta}_{RS}$ &
$10^3\ARS h/\Um^3$ \\
\midrule
\midrule
\multirow{3}{*}{1}
& 30 & $-2.1$ & $-2.8$ & $21.7$ & $22.8$ & $0.0$ & $+0.719$ & $0.55$ \\
& 45 & $+4.6$ & $+17.9$ & $15.6$ & $11.6$ & $0.0$ & $+0.802$ & $0.32$ \\
& 90 & $+6.1$ & $-4.4$ & $15.0$ & $12.8$ & $3.2$ & $+0.828$ & $0.68$ \\
\addlinespace
\midrule
\multirow{3}{*}{2}
& 30 & $+0.8$ & $+69.1$ & $68.3$ & $66.4$ & $100.0$ & $-0.709$ & $3.41$ \\
& 45 & $+6.8$ & $+72.6$ & $64.0$ & $68.9$ & $92.0$ & $-0.599$ & $1.59$ \\
& 90 & $+10.8$ & $-6.9$ & $19.7$ & $13.7$ & $10.5$ & $+0.718$ & $0.41$ \\
\addlinespace
\midrule
\multirow{3}{*}{3}
& 30 & $+0.2$ & $+67.8$ & $67.6$ & $66.8$ & $100.0$ & $-0.707$ & $8.84$ \\
& 45 & $+4.4$ & $+83.2$ & $76.1$ & $75.7$ & $98.6$ & $-0.862$ & $1.95$ \\
& 90 & $+19.5$ & $-7.9$ & $32.8$ & $16.6$ & $33.5$ & $+0.336$ & $0.18$ \\
\addlinespace
\midrule
\multirow{3}{*}{4}
& 30 & $+0.4$ & $+64.9$ & $64.6$ & $64.4$ & $100.0$ & $-0.630$ & $20.48$ \\
& 45 & $+9.3$ & $+79.1$ & $69.9$ & $68.1$ & $99.7$ & $-0.754$ & $6.12$ \\
& 90 & $+24.8$ & $-25.6$ & $56.2$ & $56.9$ & $69.0$ & $-0.277$ & $0.10$ \\
\bottomrule
\end{tabular}
\end{table*}

\subsubsection{Band-averaged production}
\label{sec:band_production}

Figure~\ref{Production_decomposition} shows the TE-face band-averaged production for all twelve configurations. The bars represent the alignment contribution $\overline{\Palign}_{\mathrm{TE}}$ and the divergence-associated contribution $\overline{\Pdiv}_{\mathrm{TE}}$, while the black diamonds show the reconstructed midplane total, $\overline{P}_{\mathrm{TE}} = \overline{\Palign}_{\mathrm{TE}}+\overline{\Pdiv}_{\mathrm{TE}}$. Positive production denotes transfer of kinetic energy from the mean flow to the turbulent fluctuations, while negative production denotes transfer in the opposite direction \citep{pope2000turbulent}. All quantities are evaluated pointwise before averaging over the baseline TE-face band. Table~\ref{tab:angle_summary} lists the corresponding principal-axis statistics, including the orientation factor $\overline{\cos(2\Delta\theta)}_{RS}$ and the stress--strain amplitude $\ARS$ that together determine $\overline{\Palign}_{\mathrm{TE}}$ through Eq.~\eqref{eq:Palign_factorized}. All quantities in Fig.~\ref{Production_decomposition} and Table~\ref{tab:angle_summary} are evaluated pointwise and then averaged over the same baseline TE rear face band. Sensitivity to the sampling region is examined in ~\ref{app:sampling_robustness}, while the spanwise stress and midplane assumptions are examined in ~\ref{app:spanwise_robustness}.

At $L^*=1$, $\overline{\Palign}_{\mathrm{TE}}$ is negative for all three angles. The weighted misalignment remains below $45^\circ$, giving a
positive orientation factor and therefore negative alignment production. Its magnitude is small because the stress--strain amplitude $\ARS$ is also
small.

A marked change occurs at $L^*=2$. For the $30^\circ$ and $45^\circ$ faces, the weighted misalignment exceeds $45^\circ$ and the orientation factor changes sign, giving positive $\overline{\Palign}_{\mathrm{TE}}$. The amplitude-weighted fraction of the band with local misalignment $\Delta\theta>45^\circ$, denoted $F_{>45}$, exceeds $90\%$ for both cases. This reorganization occurs at $L^*=2$, while the reverse-flow topology still exhibits a single connected WSR. Distinct SSR and WSR regions appear only for the longer bodies. Principal-axis misalignment has similarly been used to characterize departures from linear stress--strain relations in transitional and separated turbulent flows
\citep{schmitt2007boussinesq,lengani2020principal,corsini2025transport}.

Once the orientation favors positive alignment production, the increase in $\overline{\Palign}_{\mathrm{TE}}$ with $L^*$ is governed primarily by the stress--strain amplitude. For the $30^\circ$ face, $\overline{\cos(2\Delta\theta)}_{RS}$ remains close to $-0.7$ from $L^*=2$ to $4$, whereas $10^3\ARS h/\Um^3$ increases from $3.41$ to $20.48$, as shown in Table~\ref{tab:angle_summary}. Over the same range, $\overline{\Palign}_{\mathrm{TE}}h/\Um^3$ increases from approximately $2.4\times10^{-3}$ to $12.9\times10^{-3}$. The $45^\circ$ face follows the same trend at lower amplitude. At $L^*=3$, for example, its orientation factor is more negative than that of the $30^\circ$ case ($-0.862$ compared with $-0.707$), but its stress--strain amplitude is less than one quarter as large. Its alignment production is therefore substantially smaller. The largest band-averaged alignment contribution occurs at $(L^*,\alpha)=(4,30^\circ)$.

The vertical face behaves differently. Its stress--strain amplitude decreases with increasing $L^*$, as $10^3\ARS h/\Um^3$ falls from $0.68$ at $L^*=1$ to $0.10$ at $L^*=4$. The orientation factor becomes negative at $L^*=4$, but the amplitude is too small to produce an appreciable positive alignment contribution. Consequently, $\overline{\Palign}_{\mathrm{TE}}$ remains small for the longer $90^\circ$ cases. Their negative reconstructed total production is dominated by $\overline{\Pdiv}_{\mathrm{TE}}$. As discussed in Sec.~\ref{sec:align_div}, this contribution completes the reconstructed midplane production budget but is not interpreted as an independent production mechanism.

These results separate the roles of orientation and amplitude in the near-face energy exchange. The relative principal-axis orientation controls the sign of $\overline{\Palign}_{\mathrm{TE}}$, with the inclined faces transitioning to positive alignment production by $L^*=2$. The subsequent growth with $L^*$ is driven primarily by the increasing stress--strain amplitude. This growth is strongest for the shallow face, where the region of strong production in the separated shear layer remains close to the rear surface and increasingly interacts with the flow sampled by the TE rear face band.

\subsubsection{Spatial structure of the near-face production}
\label{sec:spatial_production}

\begin{figure}[t]
\centering
\includegraphics[width=0.47\textwidth]
{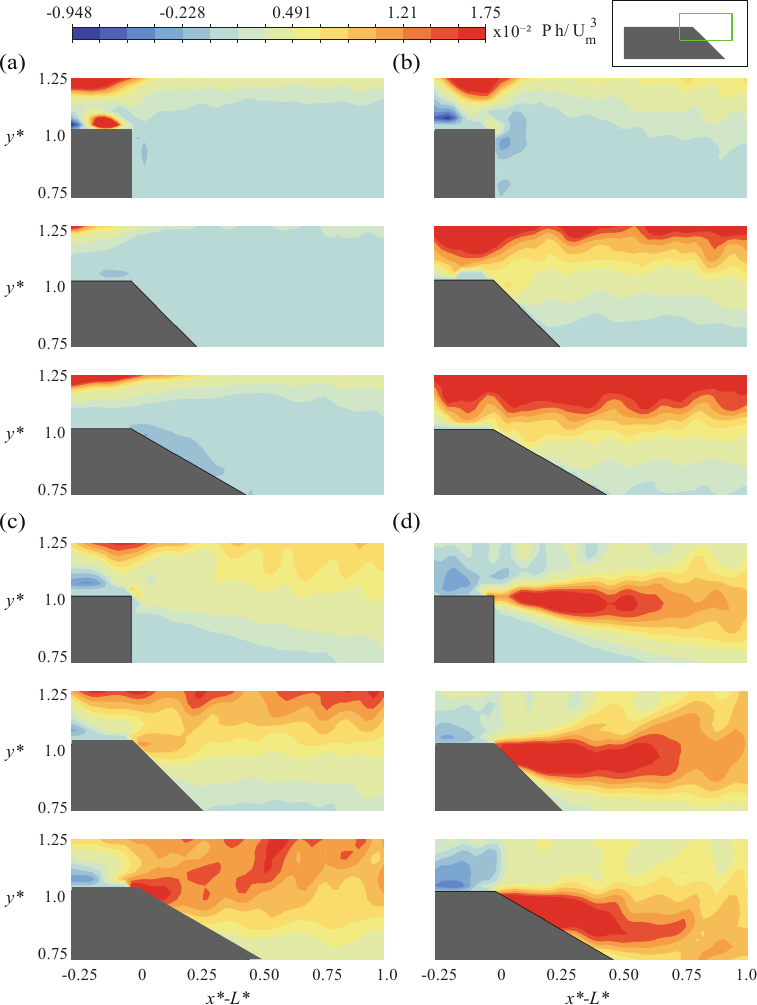}
\caption{Signed reconstructed midplane turbulent kinetic-energy production, 
$Ph/\Um^3$, in the immediate TE region, shown using a common color scale.
Panels (a)--(d) correspond to $L^*=1$--4. Within each panel,
$\alpha=90^\circ$, $45^\circ$, and $30^\circ$ are shown from top to bottom.
The quantitative averages in Fig.~\ref{Production_decomposition} are
evaluated only within the geometry-following TE-face band.}
\label{fig:production_contours}
\end{figure}

Figure~\ref{fig:production_contours} shows the signed reconstructed midplane production, $Ph/\Um^3$, near the trailing edge for all twelve configurations on a common color scale. Unlike the band averages in Fig.~\ref{Production_decomposition}, the spatial fields show where positive and negative production occur relative to the rear face and the separated flow. Production also occurs farther from the body, but this region lies outside the geometry-following band used for the quantitative near-face analysis.

At $L^*=1$, production close to the rear face is weak for all three angles. A region of relatively high production occurs upstream of the TE for $\alpha=90^{\circ}$, as expected from the sharp local turn of the return flow. At $L^*=2$, a distinct positive-production region develops near the inclined rear faces. This region strengthens with increasing $L^*$ and, for the $30^\circ$ cases, extends progressively along the shallow face. The spatial growth is consistent with the increase in the stress--strain amplitude $\ARS$ reported in Table~\ref{tab:angle_summary}.

The vertical face shows a different spatial organization. The region of strong positive production extends downstream from the TE corner and away from the vertical rear face, while the geometry-following sampling region remains close to the face. For the inclined cases, particularly at $30^\circ$, the positive-production region remains much closer to the rear surface and extends progressively along it as $L^*$ increases. This greater overlap with the near-face region is consistent with the much larger stress--strain amplitudes measured for the inclined cases.
 
The spatial fields therefore show how rear-face angle controls the local stress--strain interaction sampled near the surface. Along the shallow face, strong positive production remains close to the rear surface, producing large stress and strain amplitudes in the adjacent near-face flow. When the relative principal-axis orientation favors positive production, the larger amplitude yields the strong near-face exchange observed in Fig.~\ref{Production_decomposition}. For the vertical face, the energetic production region is displaced downstream from the surface and the stress--strain amplitude within the geometry-following band remains small. The angle dependence of the near-face production therefore reflects both the relative orientation of the stress and strain tensors and the magnitude of their local interaction.
\subsection{Global vs local effects of geometry}
\label{sec:geometric_control}

\begin{figure*}[t]
\centering
\includegraphics[width=0.90\textwidth]
{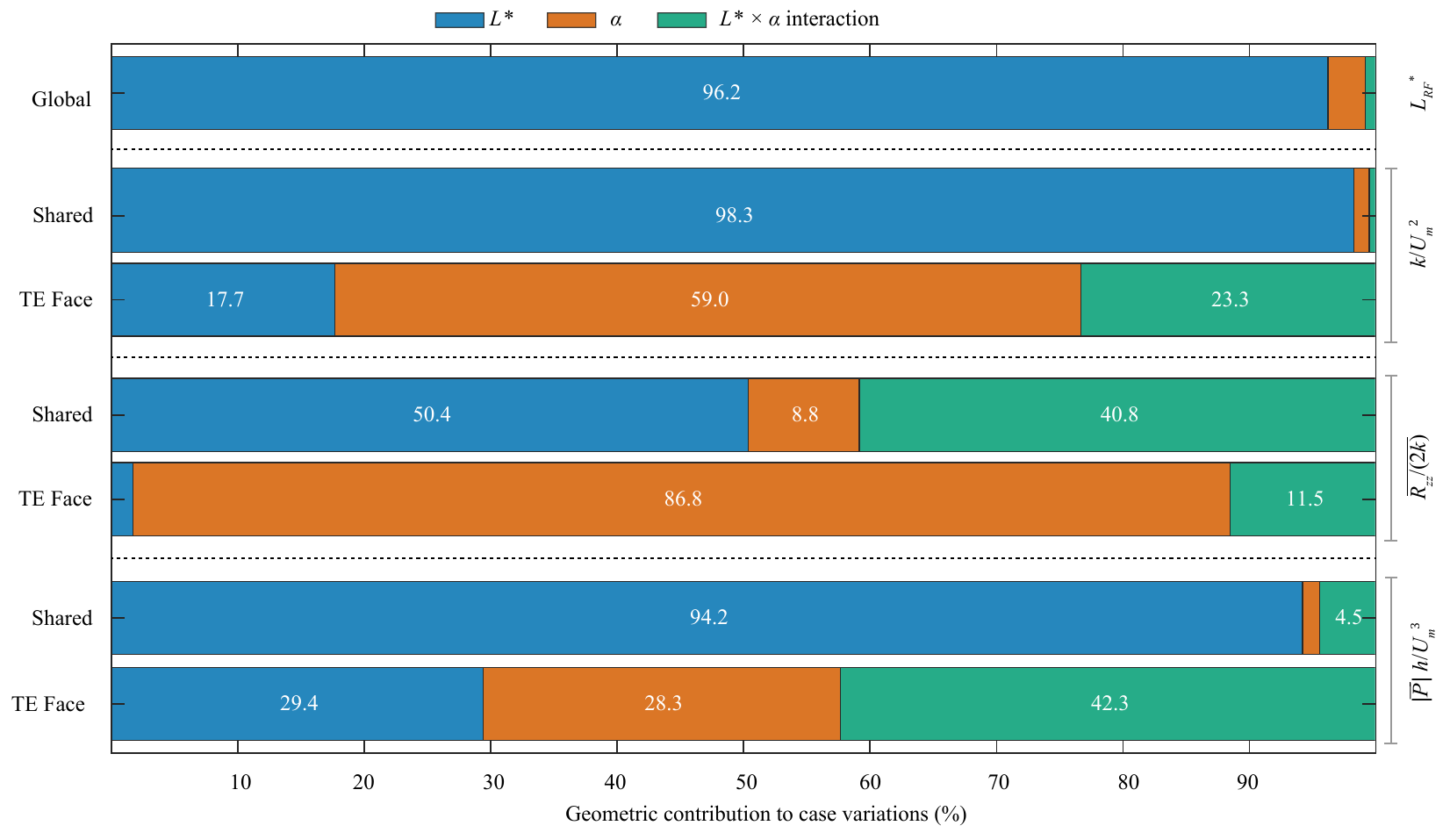}
\caption{Descriptive decomposition of the case-to-case variation into
contributions associated with protrusion length $L^*$, trailing-edge angle
$\alpha$, and their interaction. WSR reverse-flow extent $L_{\mathrm{RF}}^*$, geometry-independent shared-fluid quantities, and geometry-following TE-face quantities are shown. The percentages describe the balanced $4\times3$ geometry matrix and are not
statistical significance measures.}
\label{sensitivity}
\end{figure*}

Figure~\ref{sensitivity} summarizes the contributions of $L^*$ and $\alpha$ to the measured case-to-case variation. The twelve cases form a complete $4\times3$ matrix, with one measured case for each combination of $L^*$ and $\alpha$. Because the matrix is complete and balanced, the total case-to-case variation of each scalar quantity $Y_{ij}$ can be split exactly into a length contribution, an angle contribution and an interaction term,


\begin{equation}
\begin{aligned}
\underbrace{\sum_{i=1}^{4}\sum_{j=1}^{3}
\left(Y_{ij}-\bar{Y}_{\cdot\cdot}\right)^2}_{SS_{\mathrm{total}}}
&=
\underbrace{3\sum_{i=1}^{4}
\left(\bar{Y}_{i\cdot}-\bar{Y}_{\cdot\cdot}\right)^2}_{SS_{L^*}}
\\
&\quad+
\underbrace{4\sum_{j=1}^{3}
\left(\bar{Y}_{\cdot j}-\bar{Y}_{\cdot\cdot}\right)^2}_{SS_{\alpha}}
\\
&\quad+
\underbrace{
\left(SS_{\mathrm{total}}
-SS_{L^*}
-SS_{\alpha}\right)
}_{SS_{L^*\times\alpha}} .
\end{aligned}
\label{eq:variance_split}
\end{equation}


where $\bar{Y}_{\cdot\cdot}$ is the mean over all twelve cases. The quantities $\bar{Y}_{i\cdot}$ and $\bar{Y}_{\cdot j}$ are the means at fixed length and fixed angle. The results describe the distribution of the observed variation across the geometry matrix and do not measure statistical significance. With one value per cell, there is no independent residual term. Measurement and sampling noise are confounded with the interaction, so the reported interaction shares are best viewed as upper bounds on the physical interaction. For production, the decomposition is applied to $\overline{|P|}$ and therefore describes the magnitude of near-face exchange rather than the sign reversals in Fig.~\ref{Production_decomposition}.

The decomposition is applied to
\begin{equation}
Y\in\left\{
L_{\mathrm{RF}}^*,\,
\overline{k}/\Um^2,\,
\overline{R_{zz}/(2k)},\,
\overline{|P|}h/\Um^3
\right\}.
\label{eq:geometric_metrics}
\end{equation}
namely the WSR reverse-flow extent, turbulent kinetic energy, spanwise energy fraction and production activity. The global measure is the WSR extent $L_{\mathrm{RF}}^*$ listed in Table~\ref{tab:reattachment_lengths}, which describes the streamwise extent of the connected wake reverse-flow region. The remaining quantities are evaluated over two sampling regions chosen to distinguish global from local geometric effects. Production activity is evaluated using $P=P_{2D}+P_{zz}$, consistent with the midplane production defined in Sec.~\ref{sec:production}. The shared-fluid region comprises the spatial points that remain in the fluid for all three angles at a given $L^*$, so that identical physical locations are compared. The TE-face region is the geometry-following band of Sec.~\ref{sec:teface_turbulence}, which spans the same $s^*$ range for every angle and samples the flow at the same position relative to each face. The shared-fluid comparison therefore measures changes with $\alpha$ at fixed physical locations, whereas the TE-face comparison measures changes at fixed positions relative to each face. Comparing the two shows whether the observed variation is associated primarily with the large-scale separation topology or with the local rear-face geometry.

The WSR reverse-flow extent $L_{\mathrm{RF}}^*$ is controlled almost entirely by length. The same is true for the turbulent kinetic energy in the shared-fluid region. These results agree with the topology described in Sec.~\ref{sec:global_topology}. At fixed physical locations, the separation regime dominates the turbulence level.

The face-relative quantities behave differently. Near the rear face, the variation in $\overline{k}_{\mathrm{TE}}/\Um^2$ is associated mainly with $\alpha$. The spanwise fraction shows the clearest angle dependence of any measured quantity. The production activity retains a strong $L^*\times\alpha$ interaction. This is consistent with the mechanism identified above. The angle changes the position of the shear layer relative to the face, while the strength of the resulting near-face production also depends on the separation regime selected by length.

The two sampling regions answer different questions. The shared-fluid region compares the same laboratory locations across angles. The TE-face region compares the same position relative to each face. Together they show that $L^*$ controls the large-scale topology while $\alpha$ has its clearest effect in the immediate neighborhood of the rear face.

\section{Conclusions}
\label{sec:conclusions}

Stereoscopic PIV measurements over twelve trapezoidal protrusions were used to establish how upper-surface length and rear-face inclination jointly affect separation and turbulence production. The experiments covered \(L^*=1\) to \(4\) and rear-face angles of \(30^\circ\), \(45^\circ\), and \(90^\circ\), at \(Re_h\simeq5\times10^4\) with an incoming boundary-layer thickness comparable to the protrusion height. Within this range, length governs the large-scale separation topology, while inclination most strongly influences the turbulence immediately adjacent to the rear surface.

For \(L^*=1\) and \(2\), a single connected reverse-flow region extends over the upper surface and into the wake. At \(L^*=3\) and \(4\), reattachment on the upper surface produces distinct upper-surface and wake recirculation regions. The mean flow crossing the trailing-edge corner changes accordingly, from return flow for the shorter protrusions, through a near-zero mean flow close to reattachment, to predominantly forward flow at \(L^*=4\). The length is also the dominant factor in the variation of turbulent kinetic energy and production activity in the shared-fluid region. The effect of rear-face inclination therefore depends on the local flow at the trailing edge, which changes from reversed to predominantly forward as the protrusion length increases.

Inclination has a stronger influence when comparing quantities in a localized region near the rear surface. From \(L^*=2\) onward, shallower faces have higher near-face turbulent kinetic energy. At \(30^\circ\), the region of strong positive production remains close to the surface and extends progressively along it as the protrusion length increases. At the vertical face, the energetic region lies farther downstream and away from the surface, leaving weaker turbulence in the adjacent recirculating flow. For \(L^*=3\) and \(4\), the near-face turbulent kinetic energy at \(30^\circ\) is approximately five to seven times that at \(90^\circ\). The vertical face nevertheless has the largest spanwise energy fraction, showing that the magnitude and directional distribution of the turbulence respond differently to the geometry.

The decomposition of the in-plane deviatoric production distinguishes the roles of tensor orientation and magnitude in this local energy exchange. Its average over the TE-face band is weak and negative at \(L^*=1\), but becomes positive at the inclined faces at \(L^*=2\) as the weighted orientation factor changes sign. This change occurs while the upper-surface and wake reverse-flow regions remain connected, showing that the local production reorganizes before upper-surface reattachment changes the large-scale topology. The subsequent growth of the positive contribution is governed mainly by the stress-strain amplitude. At \(30^\circ\), this amplitude increases approximately sixfold between \(L^*=2\) and \(4\), while the orientation factor varies comparatively little. At the vertical face, the amplitude remains small even when the orientation favors positive production. Strong near-face production therefore requires both a relative tensor orientation that favors positive production and sufficiently large stress and strain magnitudes near the surface.

The measured orientations also identify limitations of a positive linear eddy-viscosity relation. Such a relation cannot reproduce the negative in-plane deviatoric production measured for the shortest protrusions. At the inclined faces with \(L^*\geq2\), where this contribution is positive, the weighted principal-axis misalignment remains approximately \(64^\circ\) to \(76^\circ\), below the \(90^\circ\) required by the linear relation. Predicting the correct sign of production therefore does not establish that a model captures the measured stress-strain interaction.

These findings show why the extent of separation alone is insufficient to assess the effect of a surface protrusion. A shallower rear face can shorten the wake reverse-flow region while sustaining stronger turbulence and production close to the surface. This distinction matters when selecting geometries intended to influence near-wall mixing, because the location and strength of turbulent energy exchange depend on both the flow reaching the trailing edge and the position of the shear layer relative to the rear face. The measured energy partition and production decomposition provide further tests of turbulence models beyond agreement with the mean separation pattern.

\section*{Declaration of competing interest}
The authors declare that they have no known competing financial interests or personal relationships that could have appeared to influence the work reported in this paper.

\section*{Data availability}
Data will be made available on request.



\appendix

\section{Validation and robustness checks}
\label{app:validation}

This appendix summarizes checks most directly related to the
geometry-following TE- rear face analysis, the spanwise stress and midplane production estimate, and the reattachment locations used in the main text.


\subsection{TE-face sampling and geometric robustness}
\label{app:sampling_robustness}

The TE-face analysis of Sec.~\ref{sec:teface_turbulence} samples the flow
close to the rear face over the common range
$0.05\leq s^*\leq0.90$. Sensitivity to the sampling region was examined using 27 nearby definitions with normal locations between
$2\Delta$ and $4\Delta$, upper-corner trims of $1$--$3\Delta$, and
lower-corner trims of $3$--$5\Delta$. At each $L^*$, the same valid
$s^*$ locations were retained for all three values of $\alpha$.

Across the twelve configurations, the median of the maximum deviations from the baseline sampling region in each case was $8.3\%$ for $\overline{k}/\Um^2$, $4.8\%$ for
$\overline{R_{zz}/(2k)}$, and $26.7\%$ for
$\overline{|P|}h/\Um^3$. The corresponding worst-case
deviations were $16.0\%$, $7.4\%$, and $45.7\%$, respectively. The greater sensitivity of production is consistent with its dependence on spatial velocity gradients.

The geometric decomposition of Fig.~\ref{sensitivity} was repeated over the
same support definitions. The $L^*$ contribution to shared-fluid
$\overline{k}/\Um^2$ remained between $97.2\%$ and $98.3\%$, while the $\alpha$
contribution to TE-face $\overline{k}_{\mathrm{TE}}/\Um^2$ remained between $57.7\%$ and $59.9\%$.
For the TE-face spanwise fraction, the $\alpha$ contribution remained between
$84.9\%$ and $90.3\%$. The TE-face production activity retained a substantial
$L^*\times\alpha$ interaction contribution of $35.9$--$49.2\%$. 
The global-versus-local ordering identified in the main text is therefore preserved across the tested sampling regions.


\subsection{Spanwise stress and midplane production}
\label{app:spanwise_robustness}

The out-of-plane velocity component is more sensitive to stereo-PIV reconstruction uncertainty than the in-plane components. The robustness of the measured spanwise stress was therefore examined in the outer-flow region defined by $x^*\geq0$ and $y^*\geq1.8$. Writing $q_{zz}=R_{zz}/\Um^2$, a conservative background-variance level was defined from the median and median absolute deviation as
\begin{equation}
\begin{aligned}
\sigma_{\mathrm{MAD}}
&=
1.4826\,\operatorname{MAD}(q_{zz}),\\
R_{zz,\mathrm{floor}}^*
&=
\operatorname{median}(q_{zz})
+
2\sigma_{\mathrm{MAD}} .
\end{aligned}
\label{eq:app_variance_floor}
\end{equation}
The corresponding dimensional level was subtracted from the measured
$R_{zz}$ before recalculating the local spanwise energy fraction.

For the $\alpha=90^\circ$ cases, this estimated background level corresponds
to $24.1$--$55.0\%$ of the measured TE-face $\overline{R_{zz}}_{\mathrm{TE}}$. Subtracting this level reduces $\overline{R_{zz}/(2k)}_{\mathrm{TE}}$ from $0.644$--$0.764$ to
$0.519$--$0.700$, which remains above the isotropic value of $1/3$.
A substantially larger uniform additive level, corresponding to
$72.3$--$84.2\%$ of the measured TE-face $R_{zz}$, would be required to reduce the corrected mean fraction to $1/3$. The elevated spanwise energy fraction measured at the vertical face therefore persists after this background-variance test.

The midplane production estimate of Sec.~\ref{sec:midplane} also relies on reflection symmetry about the nominal measurement plane. The measurable contribution that would vanish under exact symmetry is 
\begin{equation}
\begin{aligned}
P_{\mathrm{res}}
&= -\left(\langle u'w'\rangle\frac{\partial W}{\partial x}+\langle v'w'\rangle\frac{\partial W}{\partial y}\right),\\
\eta_{\mathrm{sym}}
&=\frac{\overline{|P_{\mathrm{res}}|}} {\overline{|P|}}.
\end{aligned}
\label{eq:app_symmetry}
\end{equation}
On the TE-face support,
$\eta_{\mathrm{sym}}$ ranges from $0.41\%$ to $10.29\%$ across the twelve
configurations. The largest values occur for $L^*=1$, where the production
magnitude is comparatively small.

As a separate check of the divergence-dependent contribution, the outer region $x^*\geq2$ gives
$\overline{|\Dtd|}h/\Um=0.041$--$0.056$, whereas the signed mean is only
$0.007$--$0.009$. The residual divergence therefore behaves approximately as
a zero-mean background level. Within the TE-face sampling region, the divergence exceeds
this level most clearly for the short vertical-face cases, whereas several of
the smaller $\Pdiv$ values are comparable with it. This is consistent with
the interpretation in Sec.~\ref{sec:align_div}, where $\Palign$ is used as
the primary diagnostic of the production mechanism and $\Pdiv$ provides the remaining contribution obtained using symmetry and continuity.

The model span is $B/h=2.5$, with the nominal measurement plane located
$1.25h$ from each side wall. Finite-span effects on the local
three-dimensional flow cannot therefore be excluded. The span is the same
for all twelve configurations, and the spanwise quantities are therefore used primarily to compare systematic variation with $L^*$ and $\alpha$ within the present geometry.


\subsection{Reattachment sensitivity}
\label{app:reattachment_sensitivity}

The reattachment locations in Table~\ref{tab:reattachment_lengths} are determined from the connected regions of the ensemble-mean flow satisfying $U<0$. The downstream termination of the $U=0$ contour defines the reattachment point. As an independent sensitivity check, the negative-to-positive zero crossing of the near-wall mean streamwise velocity was also evaluated at wall-normal offsets of $1\Delta$--$4\Delta$.

For the WSR, the resolved cases showed a variation in $X_r^*$ of
$0.016$--$0.373$, with a maximum absolute departure of $0.400$ from the
tabulated value. For the SSR, the corresponding variation was
$0$--$0.287$, with a maximum departure of $0.307$. These variations are
comparable with several of the smaller angle-to-angle differences at fixed
$L^*$ and support the decision not to interpret those differences as
independently resolved changes in reattachment location. They remain much
smaller than the change from a single connected reverse-flow region at
$L^*=1$--2 to separate SSR and WSR regions at $L^*=3$--4. The conclusion that $L^*$ selects the large-scale separation topology is therefore insensitive to the reattachment criterion.

\bibliographystyle{elsarticle-num-names}
\bibliography{references}

\end{document}